\documentclass[letterpaper,11pt]{article}
\pdfoutput=1
\usepackage{jheppub}
\usepackage{slashed}
\usepackage{graphicx}
\usepackage{subfigure}
\usepackage{soul}
\usepackage{amsmath}
\usepackage{multirow}
\usepackage{tablefootnote}
\usepackage{hyperref}
\usepackage{xcolor}
\usepackage{epstopdf}
\usepackage{lscape}
\usepackage{url}

\usepackage[utf8]{inputenc}

\newcommand{\beq}{\begin{equation}}
\newcommand{\eeq}{\end{equation}}
\newcommand{\bea}{\begin{eqnarray}}
\newcommand{\eea}{\end{eqnarray}}

\def\m1{M_1}
\def\m2{M_2}
\def\m3{M_3}

\newcommand{\sbeta}{s_\beta}
\newcommand{\cbeta}{c_\beta}
\newcommand{\tbeta}{t_\beta}
\newcommand{\sba}{s_{\beta-\alpha}}
\newcommand{\cba}{c_{\beta-\alpha}}

\def\ch10{\tilde \chi^0_1}

\def\gev{\,{\rm GeV}}

\def\to{\rightarrow}

\newcommand{\lsim}{\mathrel{\mathop{\kern 0pt \rlap
  {\raise.2ex\hbox{$<$}}}
  \lower.9ex\hbox{\kern-.190em $\sim$}}}
\newcommand{\gsim}{\mathrel{\mathop{\kern 0pt \rlap
  {\raise.2ex\hbox{$>$}}}
  \lower.9ex\hbox{\kern-.190em $\sim$}}}

\definecolor{pink}{RGB}{255,105,180}

\definecolor{green2}{rgb}{0,0.56,0.32}

\def\figureautorefname~#1\null{Fig.\,#1\null}
\def\tableautorefname~#1\null{Tab.\,#1\null}

\def\equationautorefname~#1\null{Eq.\,(#1)\null}

\newcommand{\tanb}{\tan \beta}
\def\cosba{\cos(\beta-\alpha)}
\def\gev{\textrm{GeV}}

\title{
Electroweak Phase Transition in 2HDM under Higgs, Z-pole, and W precision measurements}

\author[a]{Huayang Song}
\author[b]{, ~Wei Su}
\author[c]{, ~Mengchao Zhang}

\affiliation[a]{CAS Key Laboratory of Theoretical Physics, Institute of Theoretical
Physics, Chinese Academy of Sciences, Beijing 100190, P.R.China} 
\affiliation[b]{Korea Institute for Advanced Study, Seoul 02455, Korea}
\affiliation[c]{Department of Physics and Siyuan Laboratory, Jinan University, Guangzhou 510632, P.R. China}

\emailAdd{ huayangs@itp.ac.cn, weisu@kias.re.kr, mczhang@jnu.edu.cn}

\abstract{In this work we revisit the existence of a  strong  first  order electroweak phase transition (SFOEWPT) and recent $m_W$ precision measurement in the Type-I and Type-II 2HDMs. The $\mathcal{O}(100)~\gev$ new scalars in 2HDMs are favored by SFOEWPT, which is necessary for electroweak  baryogenesis, and observed $m_W$ shift as well.
We find that under current constraints, both Type-I and Type-II 2HDM can explain the SFOEWPT, Z-pole, Higgs precision measurements and $m_W$ precision measurement of CDF-II at same time, and all these precision measurements  are sensitive to heavy Higgs mass splitting in 2HDM.  The allowed regions are $\Delta m_{A/C} \in (-400, 400) ~\gev, \tanb\in(1,50)$, and $\Delta m_{A/C} \in (-200, 300) ~\gev, \tanb\in(1,12)$ for Type-I and Type-II 2HDM respectively.
Furthermore future letpon collider measurements on Higgs and $Z$ boson properties can explore this scenario in more detail or even rule out it.}

\preprint{
\begin{flushright}
KIAS--P22020
\end{flushright}
}

\begin{document}
\maketitle
\flushbottom
\section{Introduction}
With the discovery of the Higgs boson at the Large Hadron Collider (LHC) in 2012~\cite{Aad:2012tfa, Chatrchyan:2012xdj}, particle physics has been entering a new era. 
Due to the lack of direct search result at LHC, precision studies of particle physics are becoming important.
 The current measurements of particle properties seem to be consistent with all other categories of experiments and can be described by the Standard Model (SM) quite well. Meanwhile there are compelling arguments, both from theoretical and observational viewpoints, in favor of new physics beyond the Standard Model (BSM). The CDF collaboration has recently reported a precise measurement of the $W$ boson mass, which indicates a significant tension with the previous measurements and the SM prediction~\cite{CDF:2022hxs}. Although this result needs to be further confirmed by other experiments, such as D0, ATLAS, and CMS, it is still an exciting possible signal indicating the existence of new physics at a place not far above the electroweak (EW) scale~\footnote{Recent study on the new CDF result see~\cite{Fan:2022dck,Lu:2022bgw,Athron:2022qpo,Yuan:2022cpw,Strumia:2022qkt,Yang:2022gvz,deBlas:2022hdk,Zhu:2022tpr}.}.

Given this possible signal, the following question is which new physics does this possible signal point to. Among different kinds of BSM new physics, electroweak baryogenesis (EWBG)~\cite{Cline:2006ts, Morrissey:2012db} is likely to be relevant to the current possible signal. EWBG was proposed to explain the observed baryon asymmetry of the universe (BAU).
Through the baryon number breaking sphaleron process~\cite{Manton:1983nd, Klinkhamer:1984di, Kuzmin:1985mm} and CP violated scattering with bubble wall, net baryon number can be produced during the nucleation process of Higgs field.
To trigger the nucleation process and to prevent the generated net baryon being washed out, the electroweak phase transition (EWPT) needs to be a strong first order phase transition.
However, due to current measured Higgs mass, the SM EWPT is not even first order~\cite{Kajantie:1996mn,Csikor:1998eu}.

Therefore new particles are definitely required for a strong first order electroweak phase transition (SFOEWPT).
Furthermore, new particles that help to trigger SFOEWPT can not be too heavy than EW scale, otherwise they will be decoupled in the thermal phase transition process and lose effect. 
Thus, the new measure W boson mass might be a hint of EWBG. 

One method to trigger SFOEWPT is augmenting the SM Higgs sector via additional scalars, and it has been studied intensively in the literature~\cite{Carena:2018vpt,Cline:2012hg,Cline:2017qpe,Carena:2018cjh,Cline:2009sn,Moore:1998swa,Coleman:1973jx,Arnold:1992rz,Su:2020pjw,Nielsen:1975fs,Kainulainen:2019kyp,Huang:2015bta,Huang:2018aja}.
Extending the Higgs sector by a singlet scalar seems to be the simplest choice, but it is difficult for 
such model to explain the observed $m_W$ under current limits~\cite{Lopez-Val:2014jva}.  
In this work we choose Two-Higgs-Doublet Models (2HDMs)~\cite{Lee:1973iz,Branco:2011iw}
, which extend the SM Higgs sector by another doublet, as the benchmark model to study the relationship between SFOEWPT and $m_W$. 
After electroweak symmetry breaking, in addition to the SM-like Higgs boson $h$, there are three non-SM Higgs bosons, $H$/$A$/$H^{\pm}$, which can have masses below TeV and couple to the SM-like Higgs $h$ to build an energy barrier between the symmetric and broken phase. Therefore the EW phase transition can be first order and strong enough for the baryogenesis~\cite{Su:2020pjw}. Furthermore, these light extra scalar can induce a positive $m_W$ shift~\cite{Lopez-Val:2012uou} and modify the predictions on the Z-pole observables like the oblique parameters $S$, $T$ and $U$ via one-loop contributions to the $W$ and $Z$ self-energies.
The mixing between neutral scalars and extra loop corrections further reduce the Higgs couplings $\kappa_i=g_{hii}^{\rm 2HDM}/g_{hii}^{\rm SM}$ relative to their SM expectations. Though the LHC measurements still give a large amount of available phase space of the 2HDM, future Higgs factories, e.g. ILC~\cite{Bambade:2019fyw}, FCC-ee~\cite{Abada:2019lih, Abada:2019zxq} and CEPC~\cite{CEPCStudyGroup:2018ghi, CEPCPhysics-DetectorStudyGroup:2019wir} can measure them with unprecedented precision to further constrain the model.

In this paper, we study the constraints from precision measurements (especially the new $W$ boson mass and future Higgs coupling measurements) on the 2HDM and explore the possible parameter space which could lead to a SFOEWPT. 
Our study shows that SFOEWPT is consistent with the new uplifted $m_W$ in a certain parameter space.
But due to the close connection between $m_W$ and other precise measurements, the ``SFOEWPT + $m_W$'' scenario is in slight tension with current limits. 
Furthermore, the future precision lepton collider measurements of both Higgs and $Z$ boson properties could fully rule out the alive parameters to fulfil SFOEWPT and $m_W$, provided the measured central values locate in the SM prediction. 
Conversely, if the ``SFOEWPT + $m_W$'' scenario in 2HDM is true, than a clear deviation from SM prediction will be observed at future lepton colliders. 


The rest of the paper is organized as follows. In Section 2 we briefly introduce 2HDM models and related constraints. Description on EWPT is also given in Section 2. In Section 3 we perform parameter space scan on a wide range and present alive points, with future precise measurements included. 
We conclude this work in Section 4.

\section{2HDM}
\subsection{A Review}
In this section, we provide a brief review of the aspects of 2HDMs. For pedagogical introduction, see Ref.~\cite{Branco:2011iw} and a recent review Ref.~\cite{Wang:2022yhm} in light of current experiments. The scalar sector of 2HDMs consists two ${\rm SU}(2)_L$ doublets $\Phi_i$, $i=1, 2$, which can be parameterized as below,
\begin{equation}\label{eq:higgs_doublets}
    \Phi_i=\left(\begin{matrix}
        \phi_i^+\\
        (v_i+\phi_i^0+i G_i)/\sqrt{2}
    \end{matrix}\right)
\end{equation}
where $v_1$ and $v_2$ are the vacuum expectation values (VEVs) of the neutral components, satisfying the relation $v\equiv\sqrt{v_1^2+v_2^2}=246~\gev$. Assuming CP-conserving and only a soft breaking of a discrete $\mathcal{Z}_2$ symmetry allowed, the most general Higgs potential can be expressed as,
\begin{eqnarray}
\nonumber V^\text{0}(\Phi_1,\Phi_2) &=& m^2_{11}\Phi^{\dagger}_1 \Phi_1 + m^2_{22}\Phi^{\dagger}_2 \Phi_2 - m^2_{12} \left( \Phi^{\dagger}_1 \Phi_2 + h.c. \right) + \frac{\lambda_1}{2} \left( \Phi^{\dagger}_1 \Phi_1 \right)^2 + \frac{\lambda_2}{2} \left( \Phi^{\dagger}_2 \Phi_2 \right)^2 \\
& & + \lambda_3 \left( \Phi^{\dagger}_1 \Phi_1 \right) \left( \Phi^{\dagger}_2 \Phi_2 \right) + \lambda_4 \left( \Phi^{\dagger}_1 \Phi_2 \right) \left( \Phi^{\dagger}_2 \Phi_1 \right) + \frac{\lambda_5}{2} \left[ \left( \Phi^{\dagger}_1 \Phi_2 \right)^2 + h.c. \right] .
\label{potential}
\end{eqnarray}
where there are eight real parameters, $\{m_{11}^2, m_{22}^2, m_{12}^2, \lambda_1, \lambda_2, \lambda_3, \lambda_4, \lambda_5\}$. After the electroweak symmetry breaking (EWSB), the scalar sector of a 2HDM consists of five mass eigenstates: a pair of neutral CP-even Higgses, $h$ and $H$, a CP-odd Higgs, $A$, and a pair of charged Higgses $H^\pm$. We can express these states as,
\begin{equation}\begin{aligned}
\centering 
h &= - s_\alpha  \, \phi_1 + c_\alpha  \, \phi_2, 
\quad\quad\quad &A =& - s_\beta \, \varphi_1 \, + c_\beta \, \varphi_2, 
\\ H &=  \phantom{-} c_\alpha \, \phi_1 + s_\alpha \, \phi_2, 
\quad\quad\quad &H^\pm =& - s_\beta \, \phi_1^\pm + c_\beta\,
\phi_2^\pm. 
\end{aligned}\end{equation}
where we will identify $h$ with the discovered SM-like 125~\gev~Higgs without loss of generality.

For convenience, we will parametrize the potential of 2HDMs by the physical Higgs masses $m_h$, $m_H$, $m_A$ and $m_{H^\pm}$, the mixing angle between the two CP-even Higgses $\alpha$, $\tan\beta\equiv v_2/v_1$, the electroweak VEV $v$, and the soft $\mathcal{Z}_2$ symmetry breaking parameter $m_{12}^2$. Note that the vacuum expectation value $v$ and the mass of the SM-like Higgs, $m_h$ are fixed to their known values 246~\gev~and 125~\gev~respectively, leaving the remaining six independent parameters.

Assigning different $\mathcal{Z}_2$ parities to the SM fermions, there are four types of 2HDMs. However, in this study, we focus on the so-called Type-I and Type-II 2HDMs, where all fermions obtain their masses from a single Higgs doublet in Type-I model while up- and down-type fermions obtain their masses from differnt Higgs doublets in Type-II model. In the Type-II model the couplings between $A/H$ and down-type fermions are enhanced by $\tan\beta$ and therefore it is usually more constrained by experiments when $\tan\beta$ is large.


\subsection{Theoretical Constraints on 2HDMs}
The parameter spaces of 2HDMs are already constrained by theoretical consideration without experimental results.
\begin{itemize} 
\item \textbf{Vacuum stability} 
In order to make the vacuum stable, the scalar potential should be bounded from below~\cite{Gunion:2002zf}:
\begin{eqnarray}
\lambda_1 > 0 \ , \ \lambda_2 > 0 \ , \ \lambda_3 > -\sqrt{\lambda_1 \lambda_2} \ , \ \lambda_3 + \lambda_4 - |\lambda_5| > -\sqrt{\lambda_1 \lambda_2} 
\end{eqnarray}

\item \textbf{Perturbativity and unitarity}
Requiring perturbativity, we must have $|\lambda_i|\leq  4\pi$. And requiring tree-level unitarity of the scattering in the 2HDM scalar
sector imposes the following additional mass constraints~\cite{Ginzburg:2005dt}:
\begin{eqnarray}
& &\left|3(\lambda_1 + \lambda_2) \pm \sqrt{9(\lambda_1 - \lambda_2)^2 + 4(2\lambda_3 + \lambda_4)^2 }\right| < 16\pi \ , \ \\
& &\left|(\lambda_1 + \lambda_2) \pm \sqrt{(\lambda_1 - \lambda_2)^2 + 4 \lambda_4^2 }   \right| < 16\pi \ , \  \\
& &\left|(\lambda_1 + \lambda_2) \pm \sqrt{(\lambda_1 - \lambda_2)^2 + 4 \lambda_5^2 }   \right| < 16\pi \ , \ \\
& &\left|\lambda_3 + 2\lambda_4 \pm 3\lambda_5 \right| < 8\pi \ , \ \left| \lambda_3 \pm \lambda_4 \right| < 8\pi \ , \ \left| \lambda_3 \pm \lambda_5 \right| < 8\pi  
\end{eqnarray}
\end{itemize}
To understand the these constraints, it is useful to consider the relations between the quartic couplings and the physical masses
\begin{align}
v^2\lambda_1&=m_h^2-\frac{\tbeta\left(m_{12}^2-m_H^2\sbeta\cbeta\right)}{\cbeta^2}+\left(m_h^2-m_H^2\right)\left[\cba^2\left(\tbeta^2-1\right)-2\tbeta\sba\cba\right], \nonumber \\
v^2\lambda_2&=m_h^2-\frac{m_{12}^2-m_H^2\sbeta\cbeta}{\tbeta\sbeta^2}+\left(m_h^2-m_H^2\right)\left[\cba^2\left(\tbeta^{-2}-1\right)+2\tbeta^{-1}\sba\cba\right], \nonumber \\
v^2\lambda_3&=m_h^2+2m_{H^\pm}^2-2m_H^2-\frac{m_{12}^2-m_H^2\sbeta\cbeta}{\sbeta\cbeta}-\left(m_h^2-m_H^2\right)\left[2\cba^2+\sba\cba\left(\tbeta-\tbeta^{-1}\right)\right], \nonumber \\
v^2\lambda_4&=m_A^2-2m_{H^\pm}^2+m_H^2-\frac{m_{12}^2-m_H^2\sbeta\cbeta}{\sbeta\cbeta}, \nonumber \\
v^2\lambda_5&=m_H^2-m_A^2-\frac{m_{12}^2-m_H^2\sbeta\cbeta}{\sbeta\cbeta}.
\end{align}
We can introduce $\lambda v^2\equiv m_H^2-m^2_{12}/(\sbeta\cbeta)$ following Ref.~\cite{Gu:2017ckc}. The above expression indicates that the unitarity and perturbativity set up upper bounds on the mass splittings, which can be roughly taken as $\lambda v^2< 4\pi v^2$, $m_A^2-m_H^2\lesssim \mathcal{O}\left(4\pi v^2-\lambda v^2\right)$, $m_{H^\pm}^2-m_H^2\lesssim \mathcal{O}\left(4\pi v^2-\lambda v^2\right)$ and $\max\{\tbeta,\cot\beta\}\lesssim\sqrt{(8\pi v^2)/(3\lambda v^2)}$. 
Generally speaking, large mass splitting among non-SM Higgses are not allowed for large values of $\lambda v^2$ and/or non-SM Higgs masses.

\subsection{Direct searches at LEP and LHC}
The search for pair-produced charged Higgs bosons at the Large Electron-Positron Collider (LEP) imposes a lower bound of 80~\gev~on the mass of the charged Higgs boson~\cite{ALEPH:2013htx}, and LEP searches for $AH$ production constrain the sum of the masses $m_H + m_A > 209~\gev$~\cite{Schael:2006cr}.

LHC are also looking for direct productions of exotic Higgses via including $A/H \to \mu\mu$~\cite{Sirunyan:2019tkw,Aaboud:2019sgt}, $A/H \to bb$ ~\cite{Sirunyan:2018taj,Aad:2019zwb}, $A/H \to \tau\tau$~\cite{Sirunyan:2018zut,CMS:2019hvr,Aad:2020zxo}, $A/H \to \gamma\gamma$~\cite{Sirunyan:2018aui, Sirunyan:2018wnk,Aad:2014ioa,Aaboud:2017yyg, ATLAS:2018xad}, $A/H \to tt$~\cite{Sirunyan:2019wph}, $H \to ZZ$~\cite{Sirunyan:2018qlb,Aaboud:2017rel}, $H\to WW$~\cite{Sirunyan:2019pqw,Aaboud:2017gsl}, $A \to hZ \to bb\ell\ell$~\cite{Khachatryan:2015lba,Sirunyan:2019xls,Aad:2015wra,Aaboud:2017cxo}, $A \to hZ \to \tau\tau\ell\ell$~\cite{Khachatryan:2015tha,Sirunyan:2019xjg,Aad:2015wra}, $H \to hh $~\cite{Sirunyan:2017tqo,Sirunyan:2018two,Aad:2015xja,Aad:2019uzh}, and $A/H\!\to\! HZ\!/\!AZ$ \cite{Aaboud:2018eoy,Sirunyan:2019wrn}. The null results have already ruled out a significant portion of parameter space of 2HDM. For a typical mass splitting $m_A-m_H=m_{H^\pm}-m_H=200$~\gev~,
the exotic decay channel $A\to HZ$ has already excluded a neutral Higgs with mass less than $2m_t$ for $\tanb<5$ in Type-I model and for $0.5<\tanb<15$ in Type-II model. For large $\tanb$ region ($\tanb>15$) in Type-II model, this channel put the mass of neutral scalar $H$ to be above 600~\gev. Top quarks search channels, $4t$ and $A/H\to tt$, rule out $m_H<800~\gev$ for $\tanb<0.3$ and $m_H<650~\gev$ for $\tanb<1.1$ in both two types of 2HDMs. While $A/H\to\tau\tau, \gamma\gamma$ can exclude the region $m<350~\gev, \tanb<1$ in Type-I and -II models, $A/H\to\tau\tau$ could fully exclude $m_H$ larger than 800~\gev~when $\tanb>10$ in Type-II 2HDM.
For a complete recasting the LHC direct search results in the 2HDM, we refer the readers to Ref.~\cite{Kling:2020hmi, Su:2020pjw, Wang:2022yhm}.

\subsection{Z-pole and Higgs precision measurements}
Measurements of Z-pole observables at the Large Electron-Positron Collider (LEP) impose strong constraints on the 2HDM~\cite{ALEPH:2005ab}. Satisfying Z-pole constraints requires the charged scalar mass to be close to one of the heavy neutral scalar masses: $m_{H^\pm}\simeq m_H$ or $m_{H^\pm}\simeq m_A$. In our analysis, we simply take the $S, T, U$ data at $95\%$ Confidence Level (C.L.) in Tab.~\ref{tab:STU} to capture the dominant contributions from Z-pole measurements. Note that a global analysis to recast the $S, T, U$ parameters is needed by including both Z-pole observables and latest $W$ mass measurement~\cite{CDF:2022hxs}, here for simplicity we just take them as two separate measurements and are going to discuss more on the $W$ boson mass effects in Sec.~\ref{sec:mw_2HDM}. 

\begin{table}[tb]
\centering
\resizebox{\textwidth}{!}{
  \begin{tabular}{|l|c|r|r|r|c|r|r|r|c|r|r|r|c|r|r|r|c|r|r|r|}
   \hline
    & \multicolumn{4}{c|}{Current}& \multicolumn{4}{c|}{CEPC}& \multicolumn{4}{c|}{FCC-ee }&\multicolumn{4}{c|}{ILC} \\
   \hline
   \multirow{2}{*}{}
   &\multirow{2}{*}{$\sigma$} &\multicolumn{3}{c|}{correlation}
   &{$\sigma$} &\multicolumn{3}{c|}{correlation}
   &{$\sigma$} &\multicolumn{3}{c|}{correlation}
   &{$\sigma$} &\multicolumn{3}{c|}{correlation} \\
   \cline{3-5}\cline{7-9}\cline{11-13}\cline{15-17}
   &&$S$&$T$&$U$&($10^{-2}$)&$S$&$T$&$U$&($10^{-2}$)&$S$&$T$&$U$&($10^{-2}$)&$S$&$T$&$U$\\
   \hline
   $S$& $0.04 \pm 0.11$& $1$ & $0.92$ & $-0.68$ & $1.82$  & $1$     & $0.9963$       & $-0.9745$ &   $0.370$    &  $1$     &   $0.9898$    &    $-0.8394$   &   $2.57$    &   $1$    &    $0.9947$   & $-0.9431$ \\
\hline
   $T$&$0.09\pm 0.14$& $-$ & $1$ & $-0.87$ & $2.56$  &  $-$   &  $1$      &  $-0.9844$   &   $0.514$    &   $-$    &    $1$   &    $-0.8636$   &    $3.59$   &   $-$    &   $1$    &   $-0.9569$\\
\hline
   $U$& $-0.02 \pm 0.11$& $-$ & $-$ & $1$ &$1.83$  &  $-$   &  $-$     &  $1$   &   $0.416$    &   $-$    &   $-$    &    $1$   &  $2.64$     &   $-$    &   $-$    & $1$ \\
   \hline
  \end{tabular}
  }
  \caption{Estimated $S$, $T$, and $U$ ranges and correlation matrices $\rho_{ij}$  from Z-pole precision measurements of the current results~\cite{Chen:2018shg}. 
  }
\label{tab:STU}
\end{table}
\begin{figure}[ht]
  \centering
  \includegraphics[width=0.32\linewidth]{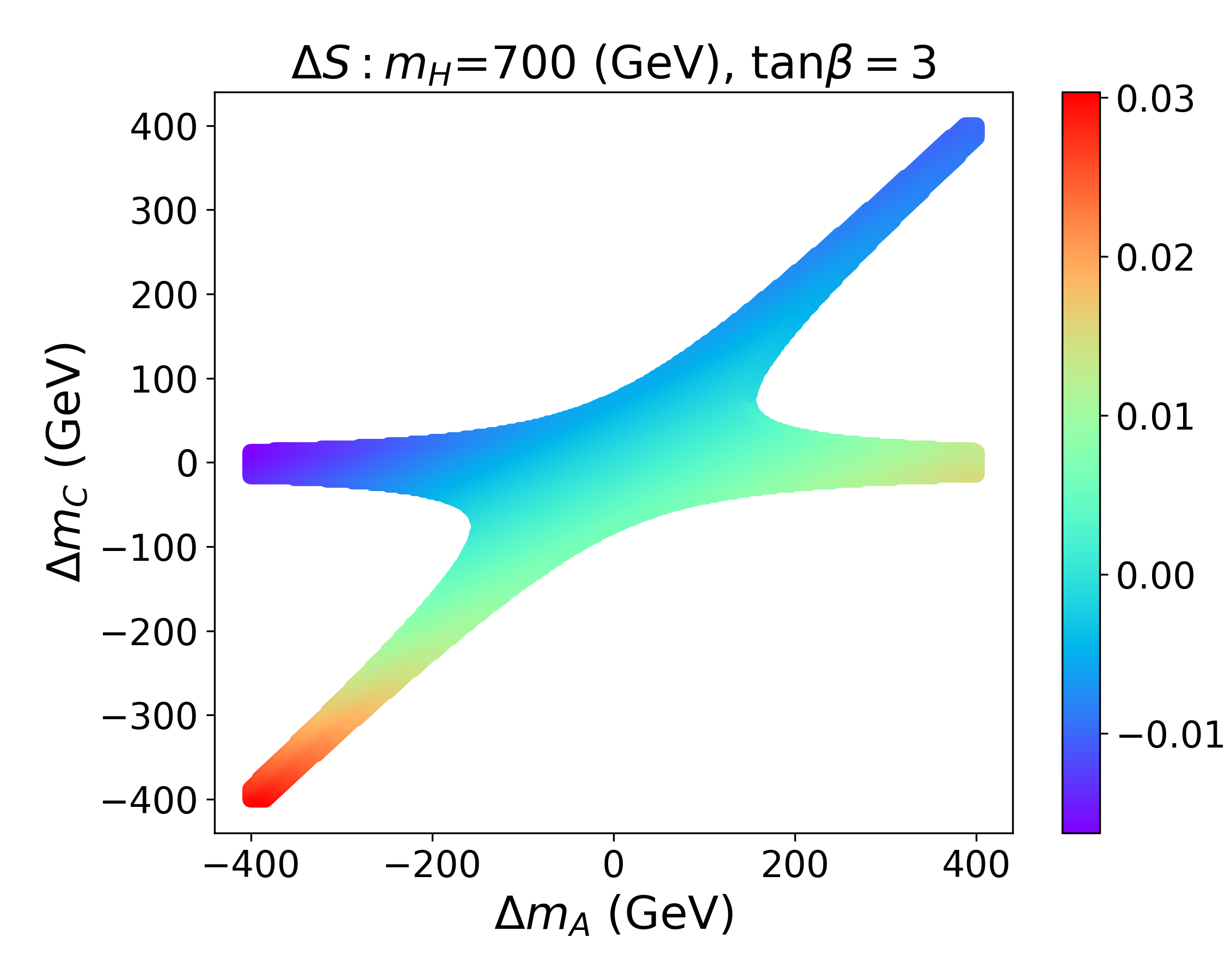}
  \includegraphics[width=0.32\linewidth]{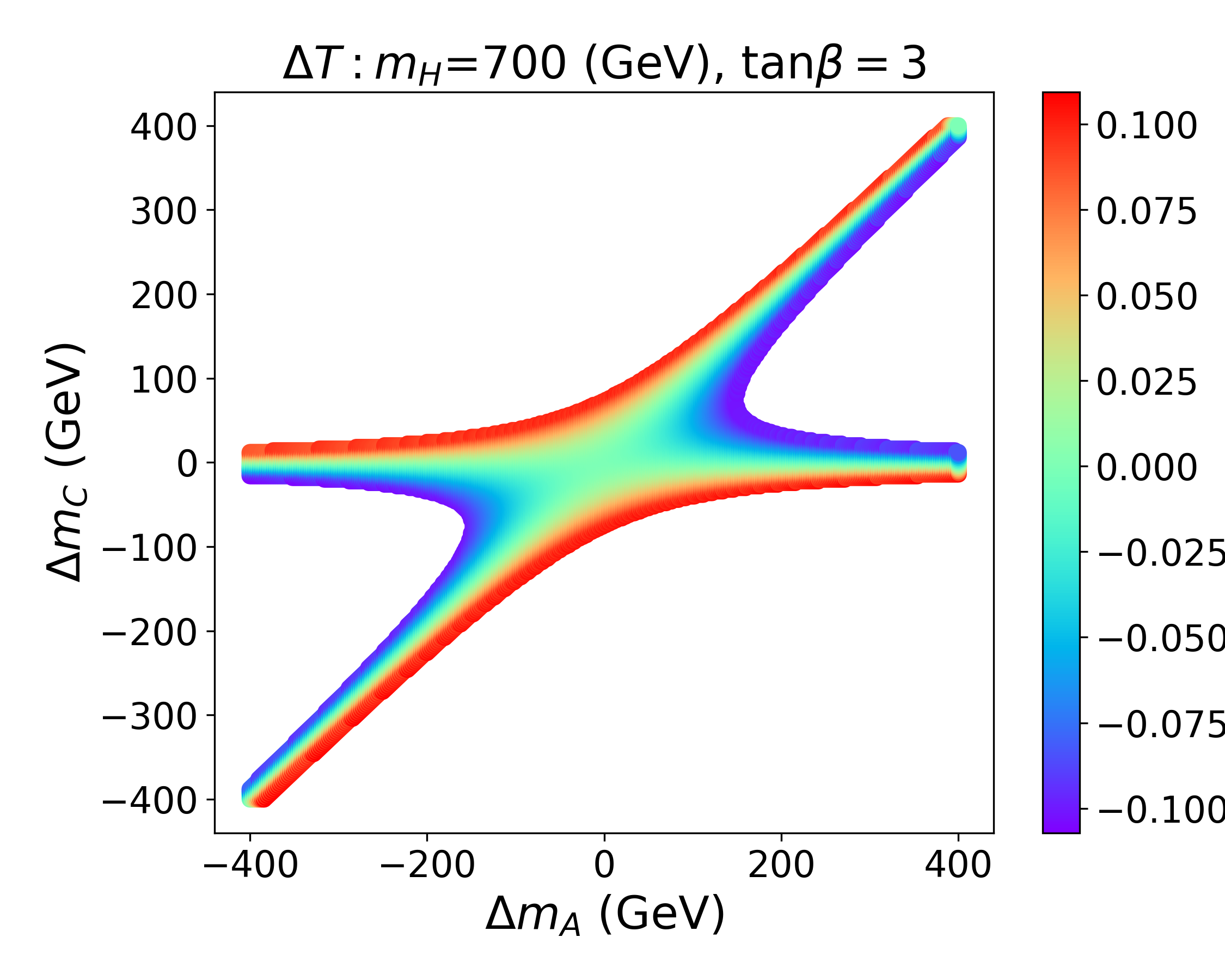}
  \includegraphics[width=0.32\linewidth]{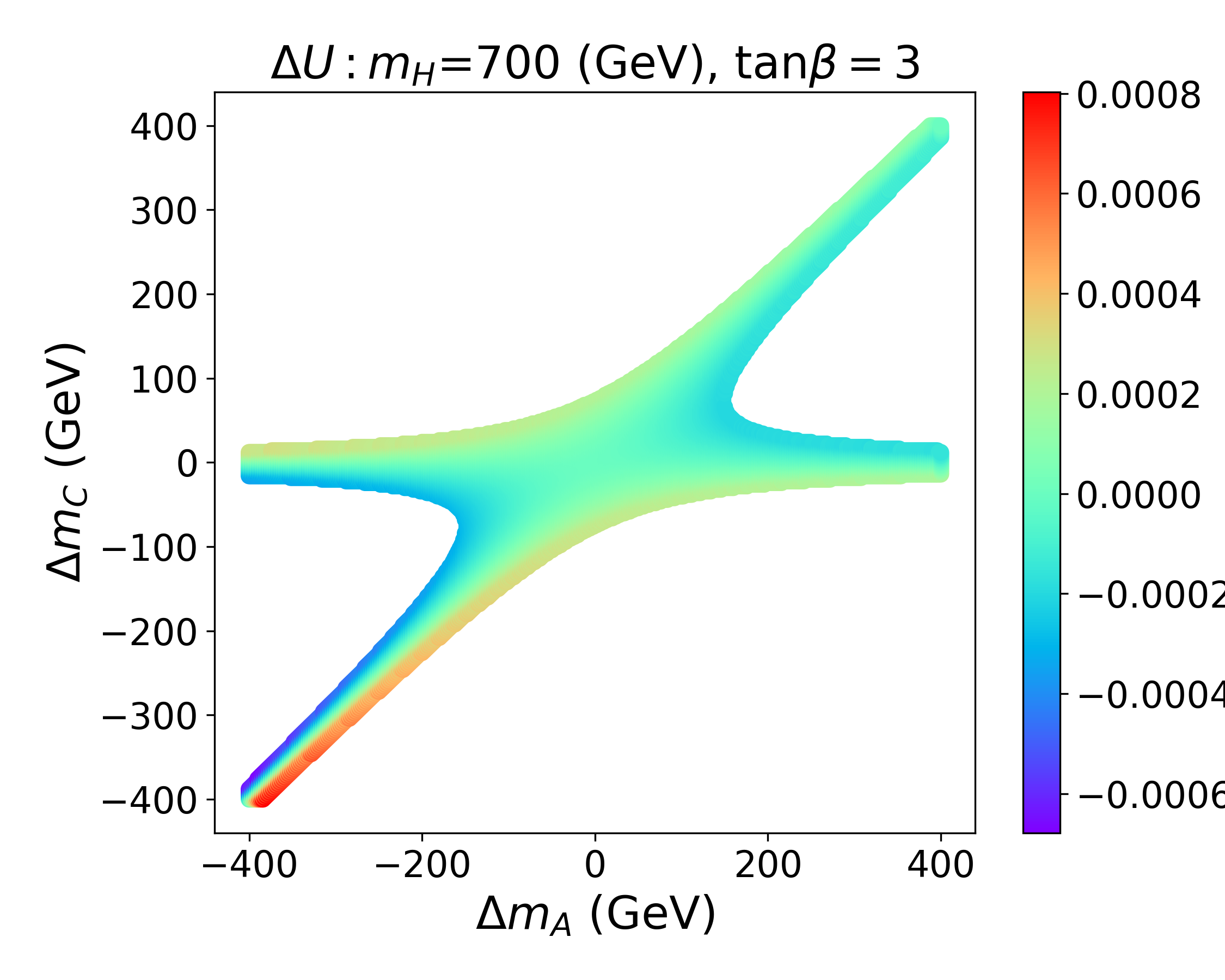}
  \caption{Current oblique constraints $S, T, U$ in the plane of  $\Delta m_A-\Delta m_C$ with $m_H=700$ GeV, $\tanb=3$ in the alignment limit $\cosba=0$. The colors are for corresponding parameter value. }
  \label{fig:t1_dmac_mH700}
\end{figure}

To reveal the relation between non-SM Higgs spectra and $S, T, U$, we define following mass splitting parameters:   
\begin{equation}
    \Delta m_A=m_A-m_H , \Delta m_C=m_{H^\pm}-m_H
\end{equation}
In~\autoref{fig:t1_dmac_mH700} we present the $S, T, U$ deviation from their SM value as functions of $\Delta m_A$ and $\Delta m_C$. 
Current measure uncertainty for $S, T, U$ are around 10\%, so it can be seen that $T$ parameter provides the most stringent limit on non-SM Higgs mass splitting. 

The LHC has also performed high precision tests on the Higgs couplings, which indicates all the measurable couplings $\kappa_i$ are close to their SM values.
Future Higgs factories will further improve the precision of measurements in the Higgs sector, and we therefore include hypothetical future lepton collider results in our study. We adopt the Higgs measurements results presented in Table 3 in Ref.~\cite{Chen:2019pkq}. Note that for future experiments, we assume there is no deviation from the SM in Higgs measurements.

\subsection{$m_W$ in the 2HDM\label{sec:mw_2HDM}}

As an important observable used in the SM precise test, $m_W$ is closely related to Z boson mass $m_Z$, Fermi constant $G_F$, and fine structure constant $\alpha$ 
\begin{eqnarray}
m^2_W \left( 1- \frac{m^2_W}{m^2_Z} \right) = \frac{\pi\alpha}{\sqrt{2}G_F} (1 + \Delta r)
\end{eqnarray}
where $\Delta r$ corresponds to quantum corrections~\cite{Lopez-Val:2012uou}.  
In the 2HDM, $m_W$ correction can be represented by~\cite{Sirlin:2012mh}
\begin{equation}
m_{W}^{\rm 2HDM}=m_{W}^{\rm SM}\left[1+\frac{\alpha {c_W}^{2}}{2\left({c_W}^{2}-{s_W}^{2}\right)} T(1+\delta\rho^{\rm 2HDM})+\frac{{\alpha}}{8 {s_W}^{2}} U-\frac{{\alpha}}{4\left({c_W}^{2}-{s_W}^{2}\right)} S\right]
\label{eq:mw_2HDM}
\end{equation}
to the $ \mathcal{O}(\alpha^2)$. Here $m_{W}^{\rm SM}=80.357 \mathrm{~GeV} \pm 4_{\text {inputs }} \pm$ $4_{\text {theory }} \mathrm{MeV}$~\cite{CDF:2022hxs}, $S, T, U$ are oblique parameters in the 2HDM~\cite{Chen:2018shg}, and $\delta\rho^{\rm 2HDM} =\frac{|\lambda_{hhh}^{\rm 2HDM}|^2}{16 \pi^2 m_h^2} $ are higher order 2HDM effects from enhanced Higgs boson self-interactions. Currently $\kappa_{hhh}=\lambda_{hhh}^{\rm 2HDM} /\lambda_{hhh}^{\rm SM} \in (-1.0,6.6)$~\cite{ATLAS:2021tyg} is already strongly constrained, thus the higher order effect $\delta\rho^{\rm 2HDM}$ up to $\mathcal{O}(0.01)$ is weak.



\begin{figure}[ht]
  \centering
     \includegraphics[width=0.48\linewidth]{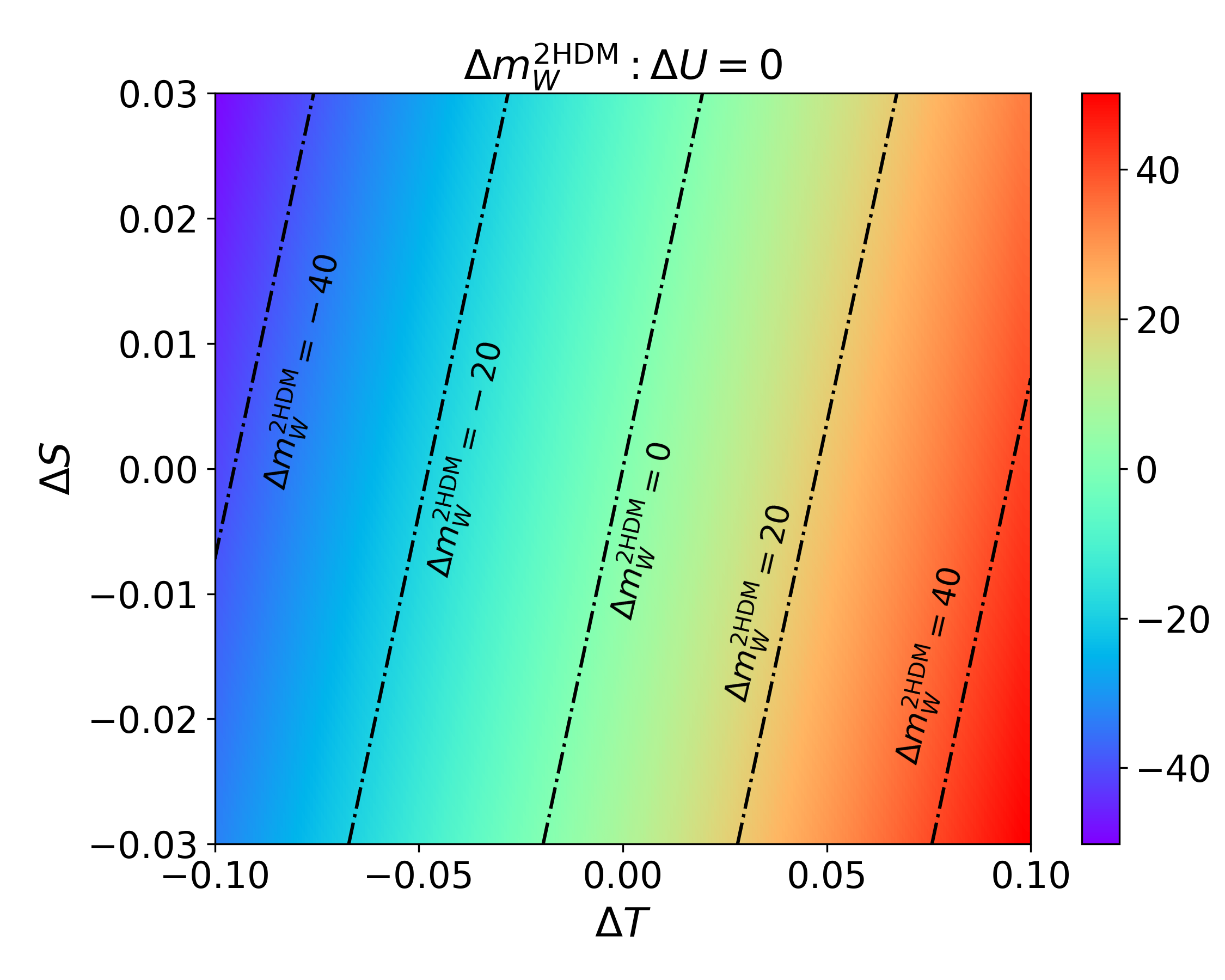}
  \includegraphics[width=0.48\linewidth]{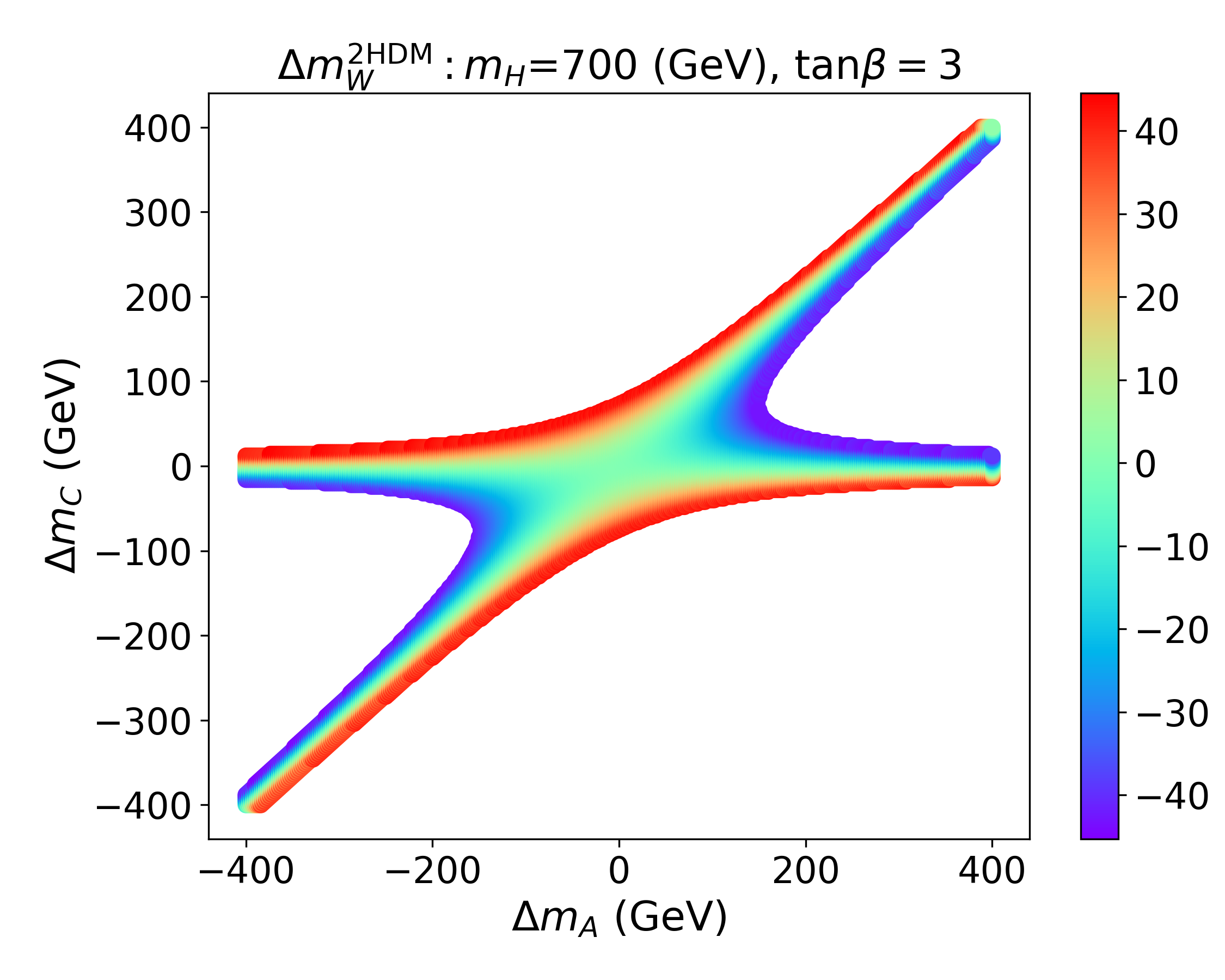}
  \caption{(Left): general picture of $\Delta m_{W}^{\rm 2HDM}$ in the plane of $\Delta T - \Delta S$ with $\Delta U=0$.  The colors are values of $\Delta m_{W}^{\rm 2HDM}$, varying from -50 MeV to 50 MeV. We have 5 black dash-dotted lines for $\Delta m_{W}^{\rm 2HDM}=-40,-20,0,20,$ and 40 MeV. (Right): $m_{W}^{\rm 2HDM}$ in the plane of $\Delta m_A - \Delta m_C$ with same benchmark spectrum $m_H=700$ GeV as~\autoref{fig:t1_dmac_mH700}. The colors are same to the left panel.}
  \label{fig:mw_BM}
\end{figure}
For convenient, we define
\begin{equation}
    \Delta m_{W}^{\rm 2HDM} = m_{W}^{\rm 2HDM}-m_{W}^{\rm SM}
\end{equation}
In the left panel of ~\autoref{fig:mw_BM} where we take $\Delta U=0$ as a benchmark case, we show the general picture of $\Delta m_{W}^{\rm 2HDM}$ in the plane of $\Delta T - \Delta S$ based on the allowed region shown in the~\autoref{fig:t1_dmac_mH700}. The colors show values of $\Delta m_{W}^{\rm 2HDM}$, varying from -50 MeV to 50 MeV. We have 5 black dash-dotted lines for $\Delta m_{W}^{\rm 2HDM}=-40,-20,0,20,$ and 40 MeV. Generally speaking, $\Delta m_{W}^{\rm 2HDM}$ mainly depends on $\Delta T$, and the larger $\Delta T$ and smaller $\Delta S$ result in large $\Delta m_{W}^{\rm 2HDM}$. This result can be easily understood with the sizes of the coefficients in front of $S, T, U$ in~\autoref{eq:mw_2HDM}.
In the right panel, we take the benchmark spectrum of~\autoref{fig:t1_dmac_mH700} with $m_H=700$ GeV and $\tanb=3$, and show $\Delta m_{W}^{\rm 2HDM}$ in the plane of $\Delta m_A - \Delta m_C$. We can see that, under current various constraints,  the benchmark spectrum here can supply for the theoretical correction meeting the new experimental measurement at CDF-II~\cite{CDF:2022hxs}.
Since in the 2HDM, $\Delta m_{W}^{\rm 2HDM}$ is directly relevant to oblique parameters, and oblique parameters further depend on non-SM Higgs mass splitting 
$\Delta m_A$ and $\Delta m_C$. 
So it is clear that $\Delta m_{W}^{\rm 2HDM}$ is sensitive to non-SM Higgs mass splitting.

\subsection{Flavour constraints}
The charged Higgs $H^{\pm}$ boson couples to both up and down type fermions, which can lead to flavor changing processes strongly constrained by flavor physics observations. The most stringent of limits comes from the measurements of B-meson decays (e.g. $b\to s\gamma$ and $B^+\to\tau\nu$), which disfavor $m_{H^{\pm}}<580$~GeV and large values of $\tanb$ respectively in Type-II 2HDM~\cite{Misiak:2017bgg, Arbey:2017gmh}, or $m_{H^{\pm}}<1$~TeV and small values of $\tanb$ ($\tan\beta < 1$) in Type-I model~\cite{Arbey:2017gmh}. However, flavor constraints can be alleviated with contributions
from other sectors of the BSM models~\cite{Han:2013mga}. In this paper, we only focus on the constraints from $B$-physics on the scalar sector.

\subsection{Phase transition in the 2HDM}

To study EWPT, we need to know the thermal effective potential, which is the free-energy density, as the function of scalar VEVs.
The thermal effective potential $V(\phi_1,\phi_2,T)$ can be schematically expressed as: 
\begin{eqnarray}
V(\phi_1,\phi_2,T) = V^\text{0}(\phi_1,\phi_2) + V^\text{CW}(\phi_1,\phi_2) + V^\text{CT}(\phi_1,\phi_2) + V^\text{T}(\phi_1,\phi_2,T) .
\label{eq:TEP}
\end{eqnarray}
Here $\phi_i$ are scalar VEVs, $T$ is the temperature of thermal system, $V^\text{0}(\phi_1,\phi_2)$ is the tree-level potential, $V^\text{CW}(\phi_1,\phi_2)$ and $V^\text{CT}(\phi_1,\phi_2)$ are Coleman-Weinberg potential and counter term respectively, and $V^\text{T}(\phi_1,\phi_2,T)$ is thermal correction. 
Detailed formulas can be found in the literature~\cite{Coleman:1973jx,Arnold:1992rz}.

In the very early universe, temperature $T$ is much higher than all the particles' mass in our model. The large effective thermal mass keep $\phi_i$ at zero and thus maintain the EW-symmetry. 
And when $T$ become much lower than EW scale, the global minimum position of $V(\phi_1,\phi_2,T)$ on $\phi_1 - \phi_2 $ plane must move to a place where $\phi^2_1 + \phi^2_2 \neq 0$ to break EW-symmetry. 
To know whether this phase transition process is first-order, we can track the minimum point with $T$ decreasing. If the minimum point (which locates in zero point when $T$ is very large) ``jump to" a non-zero point discontinuously at critical temperature $T_c$, then the EWPT should be first-order. 
This method has been numerically implemented in public package $\texttt{BSMPT}$~\cite{Basler:2018cwe}. We will use this package in this work. 

Furthermore, to prevent baryon number being washed out inside Higgs bubble, the ``wash out'' parameter~\cite{Moore:1998swa} $\xi_c \equiv {v_c}/{T_c}$ ($v_c$ is the Higgs VEV at $T_c$) should roughly be larger than 1. 
Considering the uncertainty in $\xi_c$ calculation~\cite{Nielsen:1975fs,Kainulainen:2019kyp}, we use a slightly looser criteria for SFOEWPT:
\begin{eqnarray}
\xi_c \equiv \frac{v_c}{T_c} > 0.9
\end{eqnarray}

\section{Study results}

In this section, we try to explore the SFOEWPT under various current and future constraints. Specially we have a detailed study about the latest $m_W$ result at CDF-II. 

We will firstly have a large amount of random scan points, and our study include the theoretical constraints, $B$-physics, LHC Run-II direct searches, current precision measurement of Higgs and Z-pole physics. Then the further study is performed at future Higgs factories, including CEPC, FCC-ee and ILC as shown in~\autoref{tab:STU}, to confront the SFOEWPT and $m_W$ anomaly. Our study of Higgs precision measurements will deep into one-loop level.

\subsection{Study method }

\begin{figure}[t]
  \centering
   \includegraphics[width=0.48\linewidth]{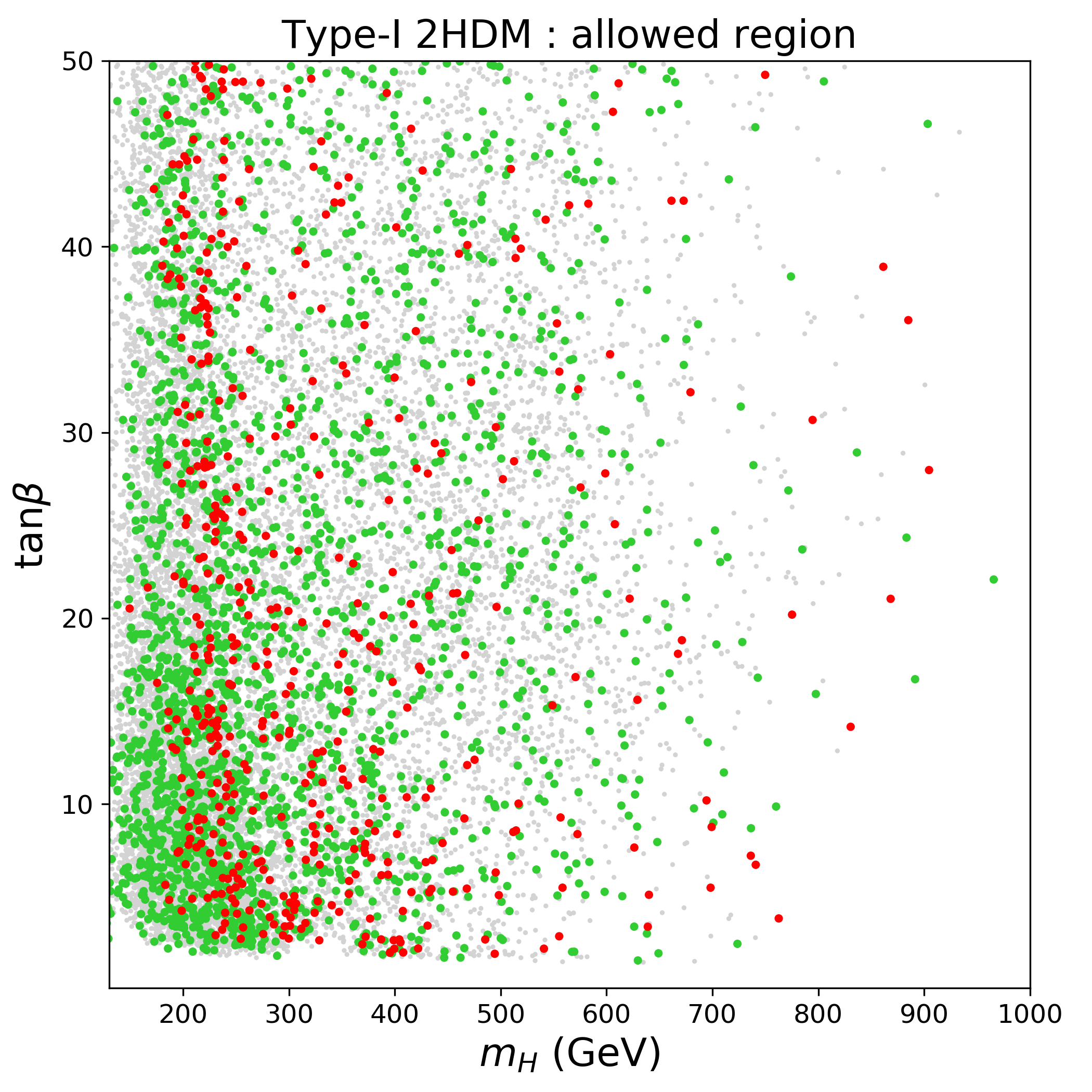}
   \includegraphics[width=0.48\linewidth]{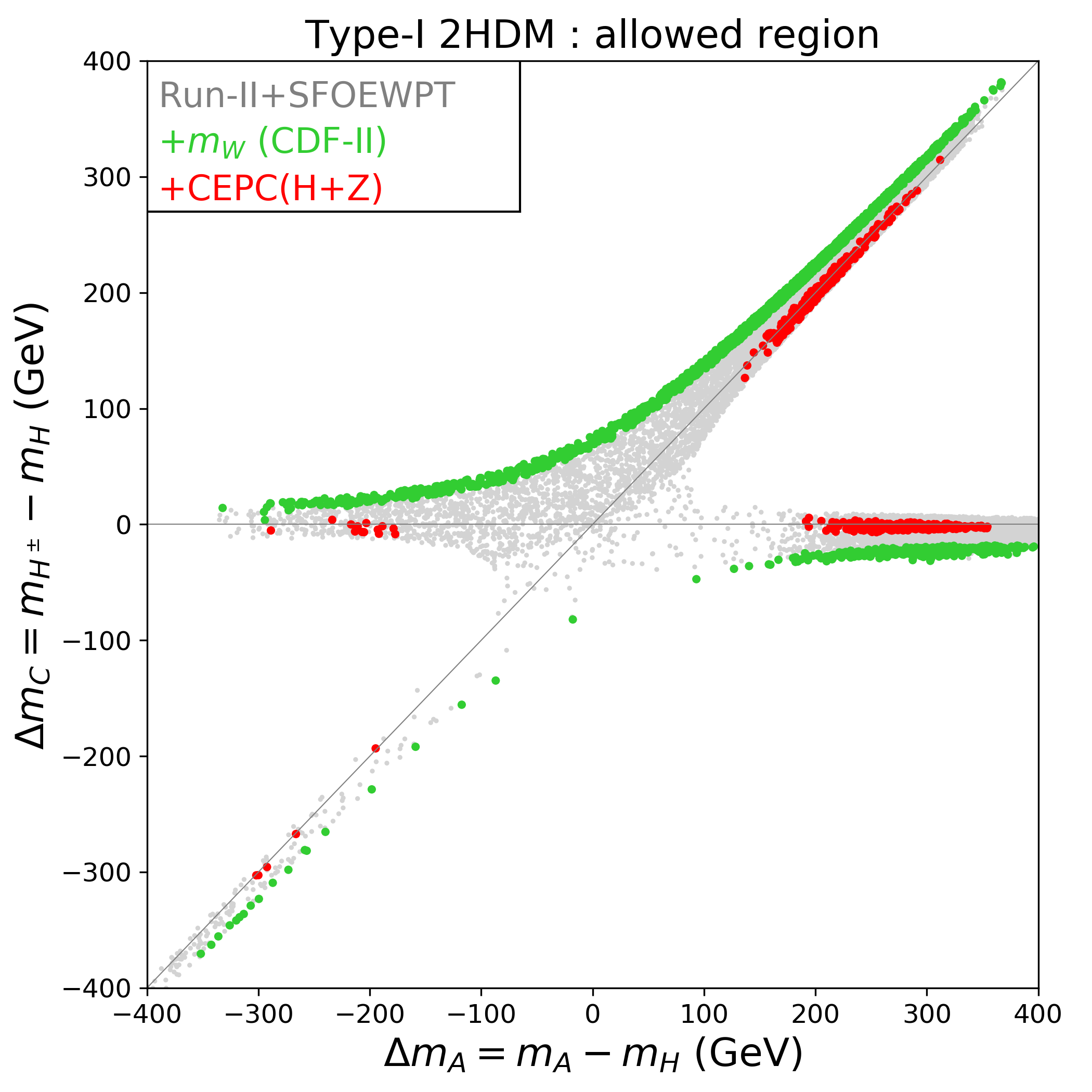} \\
  \includegraphics[width=0.48\linewidth]{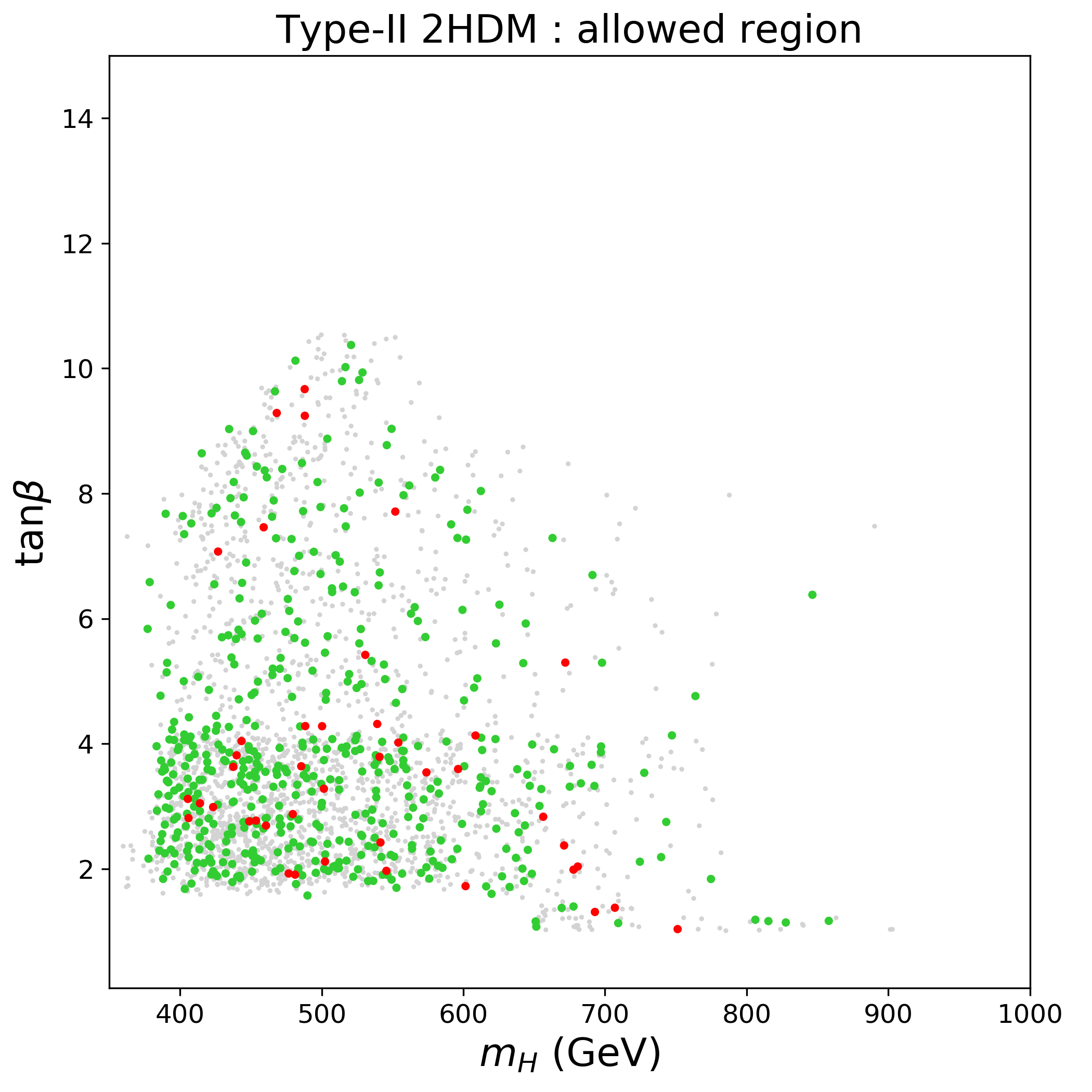}
   \includegraphics[width=0.48\linewidth]{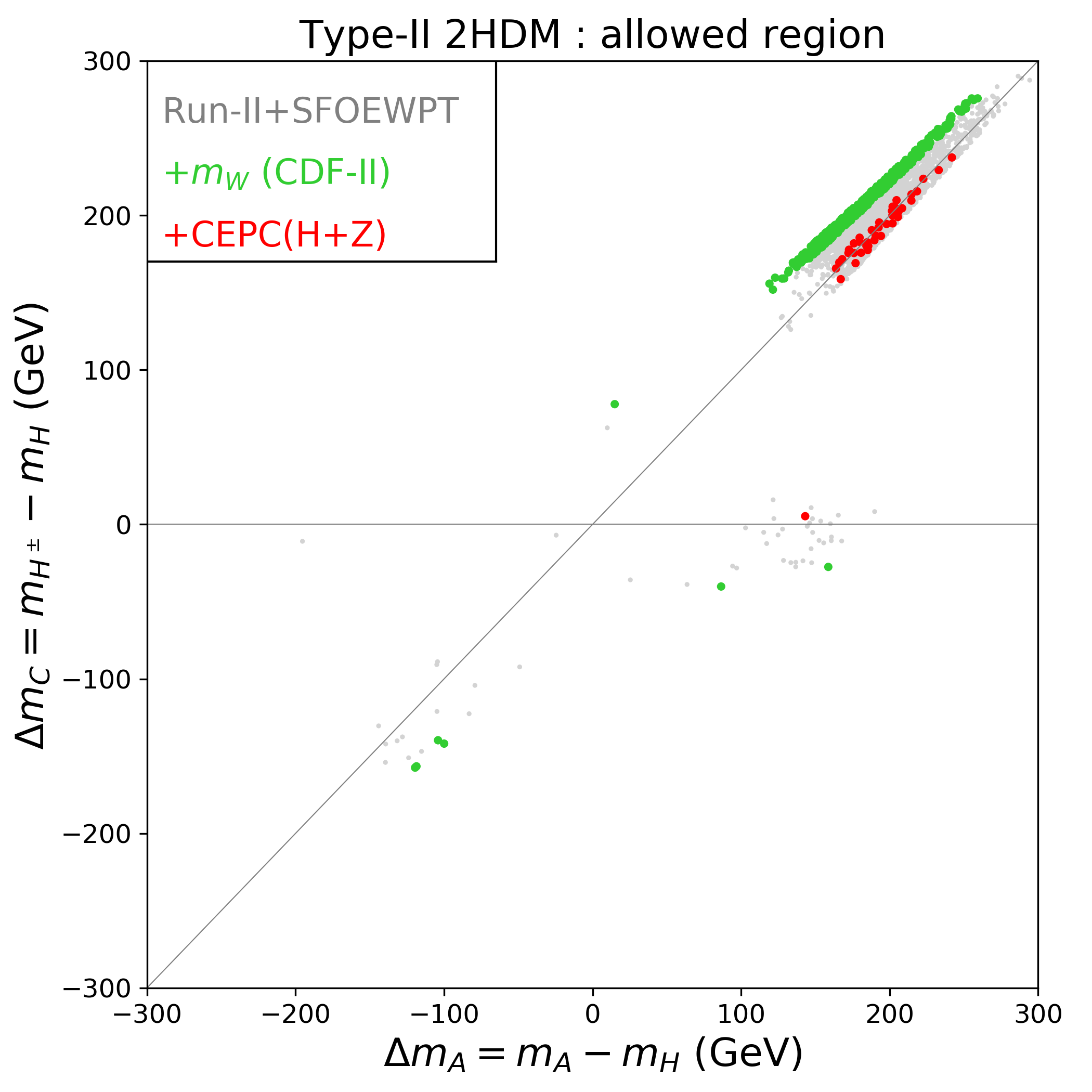}
  \caption{The allowed parameter space in the plane of $m_H - \tanb$ (left), $\Delta m_A-\Delta m_C$ (right). The grey points survive all theoretical constraints, current experimental constraints, and the conditions of SFOEWPT. The top and bottom panels are for Type-I and Type-II respectively. The green ones are able to provide a $m_W$ by CDF-II, while the red ones are  allowed by  future Higgs and Z-pole precision measurements from CEPC. The red and green points do not cover each other.}
  \label{fig:t_ma_tanb}
\end{figure}

We perform a 6 parameters random scan for both Type-I and Type-II, and the scan regions  are :
\begin{eqnarray}
& &\tan\beta \in (0.2, 50),  |\cosba|<0.5\,,  m_{A/H^{\pm}} \in (10, 1500)\text{ GeV} \ ,  \nonumber \\
& &   m^2_{12} \in(0,  1500^2) \text{ GeV}^2, \  m_H \in (130, 1500) \gev . 
\end{eqnarray}
 The number of samples allowed by current various (except for $m_W$ from CDF-II) is more than 1 million ( of 1 billion points in total) . After considering the SFOEWPT, it is about a few hundreds of thousands points allowed for Type-I, but much less for Type-II to be shown in~\autoref{fig:t_ma_tanb} as grey dots.

To incorporate in the $m_W$ at CDF-II, here in the 2HDM based on~\autoref{eq:mw_2HDM}, we take the $m_W$ data at $95\%$ Confidence Level (C.L.) with the $\chi^2$  profile-likelihood fit,
\begin{equation}
\chi^2= \frac{(m_W^{\rm{2HDM}}-m_W^{\rm{obs}})^2}{\sum \sigma_{m_W}^2}\,,
\label{eq:chiW}
\end{equation}

After taking into account of experiment uncertainties, we have
\begin{equation}
    \Delta m_{W}^{\rm 2HDM}|_{\rm ex} \in (36.3,103.4) {\rm ~MeV},
\end{equation}
and if considering SM theoretical uncertainties as well, it is $\Delta m_{W}^{\rm 2HDM}|_{\rm th+ex} \in (31.1,108.9) {\rm ~MeV}$ for the 6 parameter scan at $95\%$ C.L.. We will only take $\Delta m_{W}^{\rm 2HDM}|_{\rm ex}$ as condition of $m_W$ for CDF-II in the following study\footnote{There is no apparent difference found for $\Delta m_{W}^{\rm 2HDM}|_{\rm ex}$ and $\Delta m_{W}^{\rm 2HDM}|_{\rm th+ex}$ in our study.}. 
\subsection{SFOEWPT under LHC measurements}

As discussed above, after scanning the entire parameter space of Type-I and Type-II 2HDM, we obtain the sector which is allowed by current limits and also satisfy the SFOEWPT requirement. 

As shown in~\autoref{fig:t_ma_tanb}, the grey points meet all the conditions of current various measurements(except for $m_W$ at CDF-II) and SFOEWPT inboth types. For Type-I and Type-II, favored mass region are different: 
\begin{itemize}
    \item Type-I: To satisfy SFOEWPT and current limits, $m_H$ distributes in region $(125, 1000)$ GeV with $\tanb$ varying from 1 to 50. Mass splitting $m_{H^\pm}-m_H$ and $m_{A}-m_H$ distribute in region $(-400, 400)$ GeV.   
    \item Type-II: Compared with Type-I, Type-II is more limited due to the limited $\tan\beta$ region. $m_H$ distributes in region $(125, 1000)$ GeV, while $\tanb$ is limited to $(1,12)$. Mass splitting $m_{H^\pm}-m_H$ and $m_{A}-m_H$ vary in region $(-200, 300)$ GeV. 
\end{itemize}
To summarize, SFOEWPT require the mass of non-SM Higgs $H/A/H^{\pm}$ to be smaller than 1 TeV and a certain mount of splitting between them.  
Comparing with the right panel of ~\autoref{fig:mw_BM}, it is clear that the uplifted $m_W$ is consistent with SFOEWPT requirement.

\subsection{SFOEWPT under Higgs, Z and W precision measurements}

As discussed above, SFOEWPT, Higgs precision measurements at one-loop level, and Z-pole physics (oblique parameter $S, T, U$) are all connected by heavy Higgs mass splitting. In more detail, \autoref{fig:mw_BM} and \autoref{eq:mw_2HDM} tells that non-zero $\Delta S/T$ is needed for uplifting $m_W$. But Higgs and Z-pole physics have strong constraints on the value of $\Delta S/T$.
As presented in~\autoref{fig:t_ma_tanb}, the green points meets all these Higgs, Z-pole, $m_W$ and SFOEWPT conditions. Compared to the grey points, we can see the allowed $\tanb, m_H, \Delta m_A$ and $\Delta m_C$ region of green points does not change a lot. Another feature is they mainly locate around the boundary region, which is is mainly because current electroweak measurements is not precise enough, so the uplifted $m_W$ in need can still be satisfied within 2HDM framework. As shown in our benchmark case~\autoref{fig:t1_dmac_mH700}, specific $\Delta T$ is in need. Since oblique parameters is type universal, thus Type-I and Type-II have similar features for green region.

However, future lepton colliders, such as CEPC, ILC, and FCC-ee, will measure electroweak parameters to unprecedented precision. As presented in Tab.~\ref{tab:STU}, uncertainties of oblique measurements in CEPC can be reduced to 1\% level, which one order smaller than current uncertainties. For the Higgs precision measurements, the works~\cite{Gu:2017ckc,Chen:2019pkq,Chen:2018shg} have discussed them for the case of 2HDM systematically. 
Provided that there is no apparent deviation of Higgs and Z-pole proprieties to the SM predictions observed, or in other words, the future measurements turn out to be consistent with SM prediction,  we take CEPC precision measurements as an example to study the impact from future lepton colliders. Finally as shown in~\autoref{fig:t_ma_tanb}, the red points represent spectrum meeting conditions of Higgs, Z-pole measurements, and SFOEWPT. We can see, the red region is strongly restricted to $\Delta m_A=0$ or $\Delta m_C=\Delta m_A$ for both types. For Type-I, it is $\Delta m_C=\Delta m_A, |\Delta m_A| \in (150, 350) \gev $, or $ \Delta m_C=0,
|\Delta m_A| \in (150, 350)\gev$. While for Type-II, it is  $\Delta m_C=\Delta m_A$ and $\Delta m_A \in (150,250)\gev$. In both types, the red region and green region are separate from each other, which means Higgs+Z-pole measurements at CEPC can exclude the region for $m_W$ at CDF-II.

On the other hand, if SFOEWPT with uplifted $m_W$ in 2HDM is the true BSM scenario, deviations from SM prediction will be observed at future measurements with high confidence level.


\section{ Conclusion}

In this work, we revisited the existence of a  strong  first  order electroweak phase transition (SFOEWPT)  in the Type-I and Type-II 2HDMs as the grey points in~\autoref{fig:t_ma_tanb}.
At the same time, the latest precision measurement of the $m_W$ at CDF-II, indicates possible existence of new particles with mass around electroweak scale. We studied them all in the framework of 2HDM.

In detail, we carried out a global analysis, including $W$ boson mass $m_W$, SFOEWPT requirements, direct searches of scalar resonances at the LHC, and current LHC and future Higgs and Z-pole precision measurements at lepton colliders such as CEPC, ILC, FCC-ee. We found that,
\begin{itemize}
    \item[1.] Since in the 2HDM, $\Delta m_{W}^{\rm 2HDM}$ is directly relevant to oblique parameters as discussed, which is dependent on the heavy Higgs mass splitting of $\Delta m_A=m_A-m_H $and $\Delta m_C=m_{H^\pm}-m_H$, we can see $\Delta m_{W}^{\rm 2HDM}$ is sensitive to heavy Higgs mass splitting.  As a result, all these precision measurements and SFOEWPT in 2HDM are sensitive to non-SM Higgs mass splitting in 2HDM.
    \item[2.]  Under current constraints, both Type-I and Type-II 2HDM can explain  the SFOEWPT, Z-pole, Higgs precision measurements and $m_W$ precision measurement of CDF-II at same time. In the~\autoref{fig:t_ma_tanb}, we have the green points satisfying all of them, under current various constraints. Generally the allowed region are  $$m_H \in (125, 950)~\gev\ , \Delta m_{A/C} \in (-400, 400) ~\gev, \tanb\in(1,50)$$ for Type-I,
    $$m_H \in (125, 900)~\gev\ , \Delta m_{A/C} \in (-200, 300) ~\gev, \tanb\in(1,12)$$ for Type-II.
    \item[3.] With future precision measurements at CEPC, ILC, or FCC-ee, if there is no deviation to SM observed at Higgs or Z-pole physics, SFOEWPT is still allowed, but $m_W$ from CDF-II can not explained anymore by 2HDM. In other words, if 2HDM is the true BSM scenario after lepton colliders run, deviations from SM prediction will be observed at future measurements with high confidence level. \\
    $\quad~~$ Such a constrained parameter space points out a clear direction for experimental studies and also theoretical explorations for explaining other phenomenology.
\end{itemize}


\section*{Acknowledgments}
We thank Jin Min Yang, Shufang Su, Yang Zhang for useful discussions and comments. H.S. is supported by the International Postdoctoral Exchange Fellowship Program. W.S. is supported by  KIAS Individual Grant (PG084201) at Korea Institute for Advanced Study. M.Z. was supported by the National Natural Science Foundation of China (NNSFC) under Grant No. 12105118 and 11947118. 

\appendix

%

\bibliographystyle{JHEP}
\bibliography{ref_type1}

\providecommand{\href}[2]{#2}\begingroup\raggedright\begin{thebibliography}{10}

\bibitem{Aad:2012tfa}
{\bf ATLAS} Collaboration, G.~Aad et~al., {\it {Observation of a new particle
  in the search for the Standard Model Higgs boson with the ATLAS detector at
  the LHC}},  {\em Phys. Lett.} {\bf B716} (2012) 1--29,
  [\href{http://arxiv.org/abs/1207.7214}{{\tt arXiv:1207.7214}}].

\bibitem{Chatrchyan:2012xdj}
{\bf CMS} Collaboration, S.~Chatrchyan et~al., {\it {Observation of a new boson
  at a mass of 125 GeV with the CMS experiment at the LHC}},  {\em Phys. Lett.}
  {\bf B716} (2012) 30--61, [\href{http://arxiv.org/abs/1207.7235}{{\tt
  arXiv:1207.7235}}].

\bibitem{CDF:2022hxs}
{\bf CDF} Collaboration, T.~Aaltonen et~al., {\it {High-precision measurement
  of the $W$ boson mass with the CDF II detector}},  {\em Science} {\bf 376}
  (2022), no.~6589 170--176.

\bibitem{Fan:2022dck}
Y.-Z. Fan, T.-P. Tang, Y.-L.~S. Tsai, and L.~Wu, {\it {Inert Higgs Dark Matter
  for New CDF W-boson Mass and Detection Prospects}},
  \href{http://arxiv.org/abs/2204.03693}{{\tt arXiv:2204.03693}}.

\bibitem{Lu:2022bgw}
C.-T. Lu, L.~Wu, Y.~Wu, and B.~Zhu, {\it {Electroweak Precision Fit and New
  Physics in light of $W$ Boson Mass}},
  \href{http://arxiv.org/abs/2204.03796}{{\tt arXiv:2204.03796}}.

\bibitem{Athron:2022qpo}
P.~Athron, A.~Fowlie, C.-T. Lu, L.~Wu, Y.~Wu, and B.~Zhu, {\it {The $W$ boson
  Mass and Muon $g-2$: Hadronic Uncertainties or New Physics?}},
  \href{http://arxiv.org/abs/2204.03996}{{\tt arXiv:2204.03996}}.

\bibitem{Yuan:2022cpw}
G.-W. Yuan, L.~Zu, L.~Feng, and Y.-F. Cai, {\it {$W$-boson mass anomaly:
  probing the models of axion-like particle, dark photon and Chameleon dark
  energy}},  \href{http://arxiv.org/abs/2204.04183}{{\tt arXiv:2204.04183}}.

\bibitem{Strumia:2022qkt}
A.~Strumia, {\it {Interpreting electroweak precision data including the
  $W$-mass CDF anomaly}},  \href{http://arxiv.org/abs/2204.04191}{{\tt
  arXiv:2204.04191}}.

\bibitem{Yang:2022gvz}
J.~M. Yang and Y.~Zhang, {\it {Low energy SUSY confronted with new measurements
  of W-boson mass and muon g-2}},  \href{http://arxiv.org/abs/2204.04202}{{\tt
  arXiv:2204.04202}}.

\bibitem{deBlas:2022hdk}
J.~de~Blas, M.~Pierini, L.~Reina, and L.~Silvestrini, {\it {Impact of the
  recent measurements of the top-quark and W-boson masses on electroweak
  precision fits}},  \href{http://arxiv.org/abs/2204.04204}{{\tt
  arXiv:2204.04204}}.

\bibitem{Zhu:2022tpr}
C.-R. Zhu, M.-Y. Cui, Z.-Q. Xia, Z.-H. Yu, X.~Huang, Q.~Yuan, and Y.~Z. Fan,
  {\it {GeV antiproton/gamma-ray excesses and the $W$-boson mass anomaly: three
  faces of $\sim 60-70$ GeV dark matter particle?}},
  \href{http://arxiv.org/abs/2204.03767}{{\tt arXiv:2204.03767}}.

\bibitem{Cline:2006ts}
J.~M. Cline, {\it {Baryogenesis}},  in {\em {Les Houches Summer School -
  Session 86: Particle Physics and Cosmology: The Fabric of Spacetime}}, 9,
  2006.
\newblock \href{http://arxiv.org/abs/hep-ph/0609145}{{\tt hep-ph/0609145}}.

\bibitem{Morrissey:2012db}
D.~E. Morrissey and M.~J. Ramsey-Musolf, {\it {Electroweak baryogenesis}},
  {\em New J. Phys.} {\bf 14} (2012) 125003,
  [\href{http://arxiv.org/abs/1206.2942}{{\tt arXiv:1206.2942}}].

\bibitem{Manton:1983nd}
N.~S. Manton, {\it {Topology in the Weinberg-Salam Theory}},  {\em Phys. Rev.
  D} {\bf 28} (1983) 2019.

\bibitem{Klinkhamer:1984di}
F.~R. Klinkhamer and N.~S. Manton, {\it {A Saddle Point Solution in the
  Weinberg-Salam Theory}},  {\em Phys. Rev. D} {\bf 30} (1984) 2212.

\bibitem{Kuzmin:1985mm}
V.~A. Kuzmin, V.~A. Rubakov, and M.~E. Shaposhnikov, {\it {On the Anomalous
  Electroweak Baryon Number Nonconservation in the Early Universe}},  {\em
  Phys. Lett. B} {\bf 155} (1985) 36.

\bibitem{Kajantie:1996mn}
K.~Kajantie, M.~Laine, K.~Rummukainen, and M.~E. Shaposhnikov, {\it {Is there a
  hot electroweak phase transition at m(H) larger or equal to m(W)?}},  {\em
  Phys. Rev. Lett.} {\bf 77} (1996) 2887--2890,
  [\href{http://arxiv.org/abs/hep-ph/9605288}{{\tt hep-ph/9605288}}].

\bibitem{Csikor:1998eu}
F.~Csikor, Z.~Fodor, and J.~Heitger, {\it {Endpoint of the hot electroweak
  phase transition}},  {\em Phys. Rev. Lett.} {\bf 82} (1999) 21--24,
  [\href{http://arxiv.org/abs/hep-ph/9809291}{{\tt hep-ph/9809291}}].

\bibitem{Carena:2018vpt}
M.~Carena, Z.~Liu, and M.~Riembau, {\it {Probing the electroweak phase
  transition via enhanced di-Higgs boson production}},  {\em Phys. Rev. D} {\bf
  97} (2018), no.~9 095032, [\href{http://arxiv.org/abs/1801.00794}{{\tt
  arXiv:1801.00794}}].

\bibitem{Cline:2012hg}
J.~M. Cline and K.~Kainulainen, {\it {Electroweak baryogenesis and dark matter
  from a singlet Higgs}},  {\em JCAP} {\bf 01} (2013) 012,
  [\href{http://arxiv.org/abs/1210.4196}{{\tt arXiv:1210.4196}}].

\bibitem{Cline:2017qpe}
J.~M. Cline, K.~Kainulainen, and D.~Tucker-Smith, {\it {Electroweak
  baryogenesis from a dark sector}},  {\em Phys. Rev. D} {\bf 95} (2017),
  no.~11 115006, [\href{http://arxiv.org/abs/1702.08909}{{\tt
  arXiv:1702.08909}}].

\bibitem{Carena:2018cjh}
M.~Carena, M.~Quirós, and Y.~Zhang, {\it {Electroweak Baryogenesis from
  Dark-Sector CP Violation}},  {\em Phys. Rev. Lett.} {\bf 122} (2019), no.~20
  201802, [\href{http://arxiv.org/abs/1811.09719}{{\tt arXiv:1811.09719}}].

\bibitem{Cline:2009sn}
J.~M. Cline, G.~Laporte, H.~Yamashita, and S.~Kraml, {\it {Electroweak Phase
  Transition and LHC Signatures in the Singlet Majoron Model}},  {\em JHEP}
  {\bf 07} (2009) 040, [\href{http://arxiv.org/abs/0905.2559}{{\tt
  arXiv:0905.2559}}].

\bibitem{Moore:1998swa}
G.~D. Moore, {\it {Measuring the broken phase sphaleron rate
  nonperturbatively}},  {\em Phys. Rev. D} {\bf 59} (1999) 014503,
  [\href{http://arxiv.org/abs/hep-ph/9805264}{{\tt hep-ph/9805264}}].

\bibitem{Coleman:1973jx}
S.~R. Coleman and E.~J. Weinberg, {\it {Radiative Corrections as the Origin of
  Spontaneous Symmetry Breaking}},  {\em Phys. Rev. D} {\bf 7} (1973)
  1888--1910.

\bibitem{Arnold:1992rz}
P.~B. Arnold and O.~Espinosa, {\it {The Effective potential and first order
  phase transitions: Beyond leading-order}},  {\em Phys. Rev. D} {\bf 47}
  (1993) 3546, [\href{http://arxiv.org/abs/hep-ph/9212235}{{\tt
  hep-ph/9212235}}]. [Erratum: Phys.Rev.D 50, 6662 (1994)].

\bibitem{Su:2020pjw}
W.~Su, A.~G. Williams, and M.~Zhang, {\it {Strong first order electroweak phase
  transition in 2HDM confronting future Z \& Higgs factories}},  {\em JHEP}
  {\bf 04} (2021) 219, [\href{http://arxiv.org/abs/2011.04540}{{\tt
  arXiv:2011.04540}}].

\bibitem{Nielsen:1975fs}
N.~Nielsen, {\it {On the Gauge Dependence of Spontaneous Symmetry Breaking in
  Gauge Theories}},  {\em Nucl. Phys. B} {\bf 101} (1975) 173--188.

\bibitem{Kainulainen:2019kyp}
K.~Kainulainen, V.~Keus, L.~Niemi, K.~Rummukainen, T.~V. Tenkanen, and
  V.~Vaskonen, {\it {On the validity of perturbative studies of the electroweak
  phase transition in the Two Higgs Doublet model}},  {\em JHEP} {\bf 06}
  (2019) 075, [\href{http://arxiv.org/abs/1904.01329}{{\tt arXiv:1904.01329}}].

\bibitem{Huang:2015bta}
F.~P. Huang and C.~S. Li, {\it {Electroweak baryogenesis in the framework of
  the effective field theory}},  {\em Phys. Rev. D} {\bf 92} (2015), no.~7
  075014, [\href{http://arxiv.org/abs/1507.08168}{{\tt arXiv:1507.08168}}].

\bibitem{Huang:2018aja}
F.~P. Huang, Z.~Qian, and M.~Zhang, {\it {Exploring dynamical CP violation
  induced baryogenesis by gravitational waves and colliders}},  {\em Phys. Rev.
  D} {\bf 98} (2018), no.~1 015014,
  [\href{http://arxiv.org/abs/1804.06813}{{\tt arXiv:1804.06813}}].

\bibitem{Lopez-Val:2014jva}
D.~L\'opez-Val and T.~Robens, {\it {\ensuremath{\Delta}r and the W-boson mass
  in the singlet extension of the standard model}},  {\em Phys. Rev. D} {\bf
  90} (2014) 114018, [\href{http://arxiv.org/abs/1406.1043}{{\tt
  arXiv:1406.1043}}].

\bibitem{Lee:1973iz}
T.~Lee, {\it {A Theory of Spontaneous T Violation}},  {\em Phys. Rev. D} {\bf
  8} (1973) 1226--1239.

\bibitem{Branco:2011iw}
G.~C. Branco, P.~M. Ferreira, L.~Lavoura, M.~N. Rebelo, M.~Sher, and J.~P.
  Silva, {\it {Theory and phenomenology of two-Higgs-doublet models}},  {\em
  Phys. Rept.} {\bf 516} (2012) 1--102,
  [\href{http://arxiv.org/abs/1106.0034}{{\tt arXiv:1106.0034}}].

\bibitem{Lopez-Val:2012uou}
D.~Lopez-Val and J.~Sola, {\it {Delta r in the Two-Higgs-Doublet Model at full
  one loop level -- and beyond}},  {\em Eur. Phys. J. C} {\bf 73} (2013) 2393,
  [\href{http://arxiv.org/abs/1211.0311}{{\tt arXiv:1211.0311}}].

\bibitem{Bambade:2019fyw}
P.~Bambade et~al., {\it {The International Linear Collider: A Global Project}},
   \href{http://arxiv.org/abs/1903.01629}{{\tt arXiv:1903.01629}}.

\bibitem{Abada:2019lih}
{\bf FCC} Collaboration, A.~Abada et~al., {\it {FCC Physics Opportunities}:
  {Future Circular Collider Conceptual Design Report Volume 1}},  {\em Eur.
  Phys. J. C} {\bf 79} (2019), no.~6 474.

\bibitem{Abada:2019zxq}
{\bf FCC} Collaboration, A.~Abada et~al., {\it {FCC-ee: The Lepton Collider}:
  {Future Circular Collider Conceptual Design Report Volume 2}},  {\em Eur.
  Phys. J. ST} {\bf 228} (2019), no.~2 261--623.

\bibitem{CEPCStudyGroup:2018ghi}
{\bf CEPC Study Group} Collaboration, M.~Dong et~al., {\it {CEPC Conceptual
  Design Report: Volume 2 - Physics \& Detector}},
  \href{http://arxiv.org/abs/1811.10545}{{\tt arXiv:1811.10545}}.

\bibitem{CEPCPhysics-DetectorStudyGroup:2019wir}
{\bf CEPC Physics-Detector Study Group} Collaboration, {\it {The CEPC input for
  the European Strategy for Particle Physics - Physics and Detector}},
  \href{http://arxiv.org/abs/1901.03170}{{\tt arXiv:1901.03170}}.

\bibitem{Wang:2022yhm}
L.~Wang, J.~M. Yang, and Y.~Zhang, {\it {Two-Higgs-doublet models in light of
  current experiments: a brief review}},
  \href{http://arxiv.org/abs/2203.07244}{{\tt arXiv:2203.07244}}.

\bibitem{Gunion:2002zf}
J.~F. Gunion and H.~E. Haber, {\it {The CP conserving two Higgs doublet model:
  The Approach to the decoupling limit}},  {\em Phys. Rev. D} {\bf 67} (2003)
  075019, [\href{http://arxiv.org/abs/hep-ph/0207010}{{\tt hep-ph/0207010}}].

\bibitem{Ginzburg:2005dt}
I.~F. Ginzburg and I.~P. Ivanov, {\it {Tree-level unitarity constraints in the
  most general 2HDM}},  {\em Phys. Rev. D} {\bf 72} (2005) 115010,
  [\href{http://arxiv.org/abs/hep-ph/0508020}{{\tt hep-ph/0508020}}].

\bibitem{Gu:2017ckc}
J.~Gu, H.~Li, Z.~Liu, S.~Su, and W.~Su, {\it {Learning from Higgs Physics at
  Future Higgs Factories}},  {\em JHEP} {\bf 12} (2017) 153,
  [\href{http://arxiv.org/abs/1709.06103}{{\tt arXiv:1709.06103}}].

\bibitem{ALEPH:2013htx}
{\bf ALEPH, DELPHI, L3, OPAL, LEP} Collaboration, G.~Abbiendi et~al., {\it
  {Search for Charged Higgs bosons: Combined Results Using LEP Data}},  {\em
  Eur. Phys. J. C} {\bf 73} (2013) 2463,
  [\href{http://arxiv.org/abs/1301.6065}{{\tt arXiv:1301.6065}}].

\bibitem{Schael:2006cr}
{\bf ALEPH, DELPHI, L3, OPAL, LEP Working Group for Higgs Boson Searches}
  Collaboration, S.~Schael et~al., {\it {Search for neutral MSSM Higgs bosons
  at LEP}},  {\em Eur. Phys. J.} {\bf C47} (2006) 547--587,
  [\href{http://arxiv.org/abs/hep-ex/0602042}{{\tt hep-ex/0602042}}].

\bibitem{Sirunyan:2019tkw}
{\bf CMS} Collaboration, A.~M. Sirunyan et~al., {\it {Search for MSSM Higgs
  bosons decaying to $\mu^+\mu^-$ in proton-proton collisions at $\sqrt{s}=$ 13
  TeVSearch for MSSM Higgs bosons decaying to $\mu^+\mu^-$ in proton-proton
  collisions at s=13TeV}},  {\em Phys. Lett.} {\bf B798} (2019) 134992,
  [\href{http://arxiv.org/abs/1907.03152}{{\tt arXiv:1907.03152}}].

\bibitem{Aaboud:2019sgt}
{\bf ATLAS} Collaboration, M.~Aaboud et~al., {\it {Search for scalar resonances
  decaying into $\mu^{+}\mu^{-}$ in events with and without $b$-tagged jets
  produced in proton-proton collisions at $\sqrt{s}=13$ TeV with the ATLAS
  detector}},  {\em JHEP} {\bf 07} (2019) 117,
  [\href{http://arxiv.org/abs/1901.08144}{{\tt arXiv:1901.08144}}].

\bibitem{Sirunyan:2018taj}
{\bf CMS} Collaboration, A.~M. Sirunyan et~al., {\it {Search for beyond the
  standard model Higgs bosons decaying into a $\mathrm{b\overline{b}}$ pair in
  pp collisions at $\sqrt{s} =$ 13 TeV}},  {\em JHEP} {\bf 08} (2018) 113,
  [\href{http://arxiv.org/abs/1805.12191}{{\tt arXiv:1805.12191}}].

\bibitem{Aad:2019zwb}
{\bf ATLAS} Collaboration, G.~Aad et~al., {\it {Search for heavy neutral Higgs
  bosons produced in association with $b$-quarks and decaying to $b$-quarks at
  $\sqrt{s}=13$ TeV with the ATLAS detector}},
  \href{http://arxiv.org/abs/1907.02749}{{\tt arXiv:1907.02749}}.

\bibitem{Sirunyan:2018zut}
{\bf CMS} Collaboration, A.~M. Sirunyan et~al., {\it {Search for additional
  neutral MSSM Higgs bosons in the $\tau\tau$ final state in proton-proton
  collisions at $\sqrt{s}=$ 13 TeV}},  {\em JHEP} {\bf 09} (2018) 007,
  [\href{http://arxiv.org/abs/1803.06553}{{\tt arXiv:1803.06553}}].

\bibitem{CMS:2019hvr}
{\bf CMS} Collaboration, A.~M. Sirunyan et~al., {\it {Search for a low-mass
  $\tau^+\tau^-$ resonance in association with a bottom quark in proton-proton
  collisions at $\sqrt{s}=$ 13 TeV}},  {\em JHEP} {\bf 05} (2019) 210,
  [\href{http://arxiv.org/abs/1903.10228}{{\tt arXiv:1903.10228}}].

\bibitem{Aad:2020zxo}
{\bf ATLAS} Collaboration, G.~Aad et~al., {\it {Search for heavy Higgs bosons
  decaying into two tau leptons with the ATLAS detector using $pp$ collisions
  at $\sqrt{s}=13$ TeV}},  \href{http://arxiv.org/abs/2002.12223}{{\tt
  arXiv:2002.12223}}.

\bibitem{Sirunyan:2018aui}
{\bf CMS} Collaboration, A.~M. Sirunyan et~al., {\it {Search for a standard
  model-like Higgs boson in the mass range between 70 and 110 GeV in the
  diphoton final state in proton-proton collisions at $\sqrt{s}=$ 8 and 13
  TeV}},  {\em Phys. Lett.} {\bf B793} (2019) 320--347,
  [\href{http://arxiv.org/abs/1811.08459}{{\tt arXiv:1811.08459}}].

\bibitem{Sirunyan:2018wnk}
{\bf CMS} Collaboration, A.~M. Sirunyan et~al., {\it {Search for physics beyond
  the standard model in high-mass diphoton events from proton-proton collisions
  at $\sqrt{s} =$ 13 TeV}},  {\em Phys. Rev.} {\bf D98} (2018), no.~9 092001,
  [\href{http://arxiv.org/abs/1809.00327}{{\tt arXiv:1809.00327}}].

\bibitem{Aad:2014ioa}
{\bf ATLAS} Collaboration, G.~Aad et~al., {\it {Search for Scalar Diphoton
  Resonances in the Mass Range $65-600$ GeV with the ATLAS Detector in $pp$
  Collision Data at $\sqrt{s}$ = 8 $TeV$}},  {\em Phys. Rev. Lett.} {\bf 113}
  (2014), no.~17 171801, [\href{http://arxiv.org/abs/1407.6583}{{\tt
  arXiv:1407.6583}}].

\bibitem{Aaboud:2017yyg}
{\bf ATLAS} Collaboration, M.~Aaboud et~al., {\it {Search for new phenomena in
  high-mass diphoton final states using 37 fb$^{-1}$ of proton--proton
  collisions collected at $\sqrt{s}=13$ TeV with the ATLAS detector}},  {\em
  Phys. Lett.} {\bf B775} (2017) 105--125,
  [\href{http://arxiv.org/abs/1707.04147}{{\tt arXiv:1707.04147}}].

\bibitem{ATLAS:2018xad}
{\bf ATLAS} Collaboration, T.~A. collaboration, {\it {Search for resonances in
  the 65 to 110 GeV diphoton invariant mass range using 80 fb$^{-1}$ of $pp$
  collisions collected at $\sqrt{s}=13$ TeV with the ATLAS detector}}, .

\bibitem{Sirunyan:2019wph}
{\bf CMS} Collaboration, A.~M. Sirunyan et~al., {\it {Search for heavy Higgs
  bosons decaying to a top quark pair in proton-proton collisions at $\sqrt{s}
  =$ 13 TeV}},  \href{http://arxiv.org/abs/1908.01115}{{\tt arXiv:1908.01115}}.

\bibitem{Sirunyan:2018qlb}
{\bf CMS} Collaboration, A.~M. Sirunyan et~al., {\it {Search for a new scalar
  resonance decaying to a pair of Z bosons in proton-proton collisions at
  $\sqrt{s}=13 $ TeV}},  {\em JHEP} {\bf 06} (2018) 127,
  [\href{http://arxiv.org/abs/1804.01939}{{\tt arXiv:1804.01939}}]. [Erratum:
  JHEP03,128(2019)].

\bibitem{Aaboud:2017rel}
{\bf ATLAS} Collaboration, M.~Aaboud et~al., {\it {Search for heavy ZZ
  resonances in the $\ell ^+\ell ^-\ell ^+\ell ^-$ and $\ell ^+\ell ^-\nu
  \bar{\nu }$ final states using proton–proton collisions at $\sqrt{s}= 13$
  $\text {TeV}$ with the ATLAS detector}},  {\em Eur. Phys. J.} {\bf C78}
  (2018), no.~4 293, [\href{http://arxiv.org/abs/1712.06386}{{\tt
  arXiv:1712.06386}}].

\bibitem{Sirunyan:2019pqw}
{\bf CMS} Collaboration, A.~M. Sirunyan et~al., {\it {Search for a heavy Higgs
  boson decaying to a pair of W bosons in proton-proton collisions at $\sqrt{s}
  =$ 13 TeV}},  \href{http://arxiv.org/abs/1912.01594}{{\tt arXiv:1912.01594}}.

\bibitem{Aaboud:2017gsl}
{\bf ATLAS} Collaboration, M.~Aaboud et~al., {\it {Search for heavy resonances
  decaying into $WW$ in the $e\nu\mu\nu$ final state in $pp$ collisions at
  $\sqrt{s}=13$ TeV with the ATLAS detector}},  {\em Eur. Phys. J.} {\bf C78}
  (2018), no.~1 24, [\href{http://arxiv.org/abs/1710.01123}{{\tt
  arXiv:1710.01123}}].

\bibitem{Khachatryan:2015lba}
{\bf CMS} Collaboration, V.~Khachatryan et~al., {\it {Search for a pseudoscalar
  boson decaying into a Z boson and the 125 GeV Higgs boson in llbb final
  states}},  {\em Phys. Lett.} {\bf B748} (2015) 221--243,
  [\href{http://arxiv.org/abs/1504.04710}{{\tt arXiv:1504.04710}}].

\bibitem{Sirunyan:2019xls}
{\bf CMS} Collaboration, A.~M. Sirunyan et~al., {\it {Search for a heavy
  pseudoscalar boson decaying to a Z and a Higgs boson at $\sqrt{s} =$ 13
  TeV}},  {\em Eur. Phys. J.} {\bf C79} (2019), no.~7 564,
  [\href{http://arxiv.org/abs/1903.00941}{{\tt arXiv:1903.00941}}].

\bibitem{Aad:2015wra}
{\bf ATLAS} Collaboration, G.~Aad et~al., {\it {Search for a CP-odd Higgs boson
  decaying to Zh in pp collisions at $\sqrt{s} = 8$ TeV with the ATLAS
  detector}},  {\em Phys. Lett.} {\bf B744} (2015) 163--183,
  [\href{http://arxiv.org/abs/1502.04478}{{\tt arXiv:1502.04478}}].

\bibitem{Aaboud:2017cxo}
{\bf ATLAS} Collaboration, M.~Aaboud et~al., {\it {Search for heavy resonances
  decaying into a $W$ or $Z$ boson and a Higgs boson in final states with
  leptons and $b$-jets in 36 fb$^{-1}$ of $\sqrt s = 13$ TeV $pp$ collisions
  with the ATLAS detector}},  {\em JHEP} {\bf 03} (2018) 174,
  [\href{http://arxiv.org/abs/1712.06518}{{\tt arXiv:1712.06518}}]. [Erratum:
  JHEP11,051(2018)].

\bibitem{Khachatryan:2015tha}
{\bf CMS} Collaboration, V.~Khachatryan et~al., {\it {Searches for a heavy
  scalar boson H decaying to a pair of 125 GeV Higgs bosons hh or for a heavy
  pseudoscalar boson A decaying to Zh, in the final states with $h \to \tau
  \tau$}},  {\em Phys. Lett.} {\bf B755} (2016) 217--244,
  [\href{http://arxiv.org/abs/1510.01181}{{\tt arXiv:1510.01181}}].

\bibitem{Sirunyan:2019xjg}
{\bf CMS} Collaboration, A.~M. Sirunyan et~al., {\it {Search for a heavy
  pseudoscalar Higgs boson decaying into a 125 GeV Higgs boson and a Z boson in
  final states with two tau and two light leptons at $\sqrt{s}=$ 13 TeV}},
  \href{http://arxiv.org/abs/1910.11634}{{\tt arXiv:1910.11634}}.

\bibitem{Sirunyan:2017tqo}
{\bf CMS} Collaboration, A.~M. Sirunyan et~al., {\it {Search for Higgs boson
  pair production in the $bb\tau\tau$ final state in proton-proton collisions
  at $\sqrt{(}s)=8\text{ }\text{ }\mathrm{TeV}$}},  {\em Phys. Rev.} {\bf D96}
  (2017), no.~7 072004, [\href{http://arxiv.org/abs/1707.00350}{{\tt
  arXiv:1707.00350}}].

\bibitem{Sirunyan:2018two}
{\bf CMS} Collaboration, A.~M. Sirunyan et~al., {\it {Combination of searches
  for Higgs boson pair production in proton-proton collisions at $\sqrt{s} = $
  13 TeV}},  {\em Phys. Rev. Lett.} {\bf 122} (2019), no.~12 121803,
  [\href{http://arxiv.org/abs/1811.09689}{{\tt arXiv:1811.09689}}].

\bibitem{Aad:2015xja}
{\bf ATLAS} Collaboration, G.~Aad et~al., {\it {Searches for Higgs boson pair
  production in the $hh\to bb\tau\tau, \gamma\gamma WW^*, \gamma\gamma bb,
  bbbb$ channels with the ATLAS detector}},  {\em Phys. Rev.} {\bf D92} (2015)
  092004, [\href{http://arxiv.org/abs/1509.04670}{{\tt arXiv:1509.04670}}].

\bibitem{Aad:2019uzh}
{\bf ATLAS} Collaboration, G.~Aad et~al., {\it {Combination of searches for
  Higgs boson pairs in $pp$ collisions at $\sqrt{s} = $13 TeV with the ATLAS
  detector}},  \href{http://arxiv.org/abs/1906.02025}{{\tt arXiv:1906.02025}}.

\bibitem{Aaboud:2018eoy}
{\bf ATLAS} Collaboration, M.~Aaboud et~al., {\it {Search for a heavy Higgs
  boson decaying into a $Z$ boson and another heavy Higgs boson in the
  $\ell\ell bb$ final state in $pp$ collisions at $\sqrt{s}=13$ TeV with the
  ATLAS detector}},  {\em Phys. Lett.} {\bf B783} (2018) 392--414,
  [\href{http://arxiv.org/abs/1804.01126}{{\tt arXiv:1804.01126}}].

\bibitem{Sirunyan:2019wrn}
{\bf CMS} Collaboration, A.~M. Sirunyan et~al., {\it {Search for new neutral
  Higgs bosons through the H$\to$ ZA $\to \ell^{+}\ell^{-} \mathrm{b\bar{b}}$
  process in pp collisions at $\sqrt{s} =$ 13 TeV}},
  \href{http://arxiv.org/abs/1911.03781}{{\tt arXiv:1911.03781}}.

\bibitem{Kling:2020hmi}
F.~Kling, S.~Su, and W.~Su, {\it {2HDM Neutral Scalars under the LHC}},  {\em
  JHEP} {\bf 06} (2020) 163, [\href{http://arxiv.org/abs/2004.04172}{{\tt
  arXiv:2004.04172}}].

\bibitem{ALEPH:2005ab}
{\bf ALEPH, DELPHI, L3, OPAL, SLD, LEP Electroweak Working Group, SLD
  Electroweak Group, SLD Heavy Flavour Group} Collaboration, S.~Schael et~al.,
  {\it {Precision electroweak measurements on the $Z$ resonance}},  {\em Phys.
  Rept.} {\bf 427} (2006) 257--454,
  [\href{http://arxiv.org/abs/hep-ex/0509008}{{\tt hep-ex/0509008}}].

\bibitem{Chen:2018shg}
N.~Chen, T.~Han, S.~Su, W.~Su, and Y.~Wu, {\it {Type-II 2HDM under the
  Precision Measurements at the $Z$-pole and a Higgs Factory}},  {\em JHEP}
  {\bf 03} (2019) 023, [\href{http://arxiv.org/abs/1808.02037}{{\tt
  arXiv:1808.02037}}].

\bibitem{Chen:2019pkq}
N.~Chen, T.~Han, S.~Li, S.~Su, W.~Su, and Y.~Wu, {\it {Type-I 2HDM under the
  Higgs and Electroweak Precision Measurements}},  {\em JHEP} {\bf 08} (2020)
  131, [\href{http://arxiv.org/abs/1912.01431}{{\tt arXiv:1912.01431}}].

\bibitem{Sirlin:2012mh}
A.~Sirlin and A.~Ferroglia, {\it {Radiative Corrections in Precision
  Electroweak Physics: a Historical Perspective}},  {\em Rev. Mod. Phys.} {\bf
  85} (2013), no.~1 263--297, [\href{http://arxiv.org/abs/1210.5296}{{\tt
  arXiv:1210.5296}}].

\bibitem{ATLAS:2021tyg}
{\bf ATLAS} Collaboration, {\it {Combination of searches for non-resonant and
  resonant Higgs boson pair production in the $b\bar{b}\gamma\gamma$,
  $b\bar{b}\tau^{+}\tau^{-}$ and $b\bar{b}b\bar{b}$ decay channels using $pp$
  collisions at $\sqrt{s}$ = 13 TeV with the ATLAS detector}}, .

\bibitem{Misiak:2017bgg}
M.~Misiak and M.~Steinhauser, {\it {Weak radiative decays of the B meson and
  bounds on $M_{H^\pm }$ in the Two-Higgs-Doublet Model}},  {\em Eur. Phys. J.
  C} {\bf 77} (2017), no.~3 201, [\href{http://arxiv.org/abs/1702.04571}{{\tt
  arXiv:1702.04571}}].

\bibitem{Arbey:2017gmh}
A.~Arbey, F.~Mahmoudi, O.~Stal, and T.~Stefaniak, {\it {Status of the Charged
  Higgs Boson in Two Higgs Doublet Models}},  {\em Eur. Phys. J. C} {\bf 78}
  (2018), no.~3 182, [\href{http://arxiv.org/abs/1706.07414}{{\tt
  arXiv:1706.07414}}].

\bibitem{Han:2013mga}
T.~Han, T.~Li, S.~Su, and L.-T. Wang, {\it {Non-Decoupling MSSM Higgs Sector
  and Light Superpartners}},  {\em JHEP} {\bf 11} (2013) 053,
  [\href{http://arxiv.org/abs/1306.3229}{{\tt arXiv:1306.3229}}].

\bibitem{Basler:2018cwe}
P.~Basler and M.~Mühlleitner, {\it {BSMPT (Beyond the Standard Model Phase
  Transitions): A tool for the electroweak phase transition in extended Higgs
  sectors}},  {\em Comput. Phys. Commun.} {\bf 237} (2019) 62--85,
  [\href{http://arxiv.org/abs/1803.02846}{{\tt arXiv:1803.02846}}].

\end{thebibliography}\endgroup

\end{document}